Figure 1

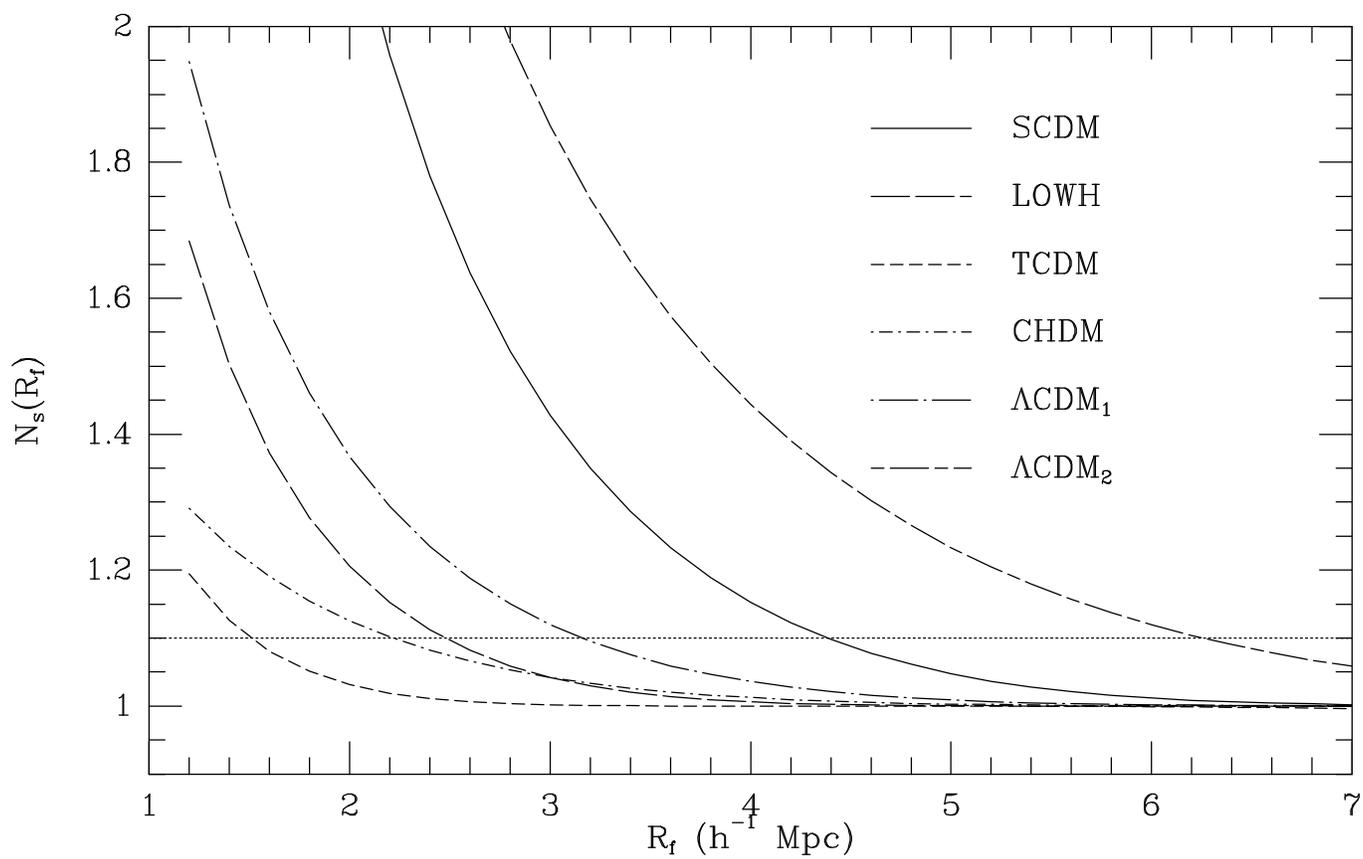

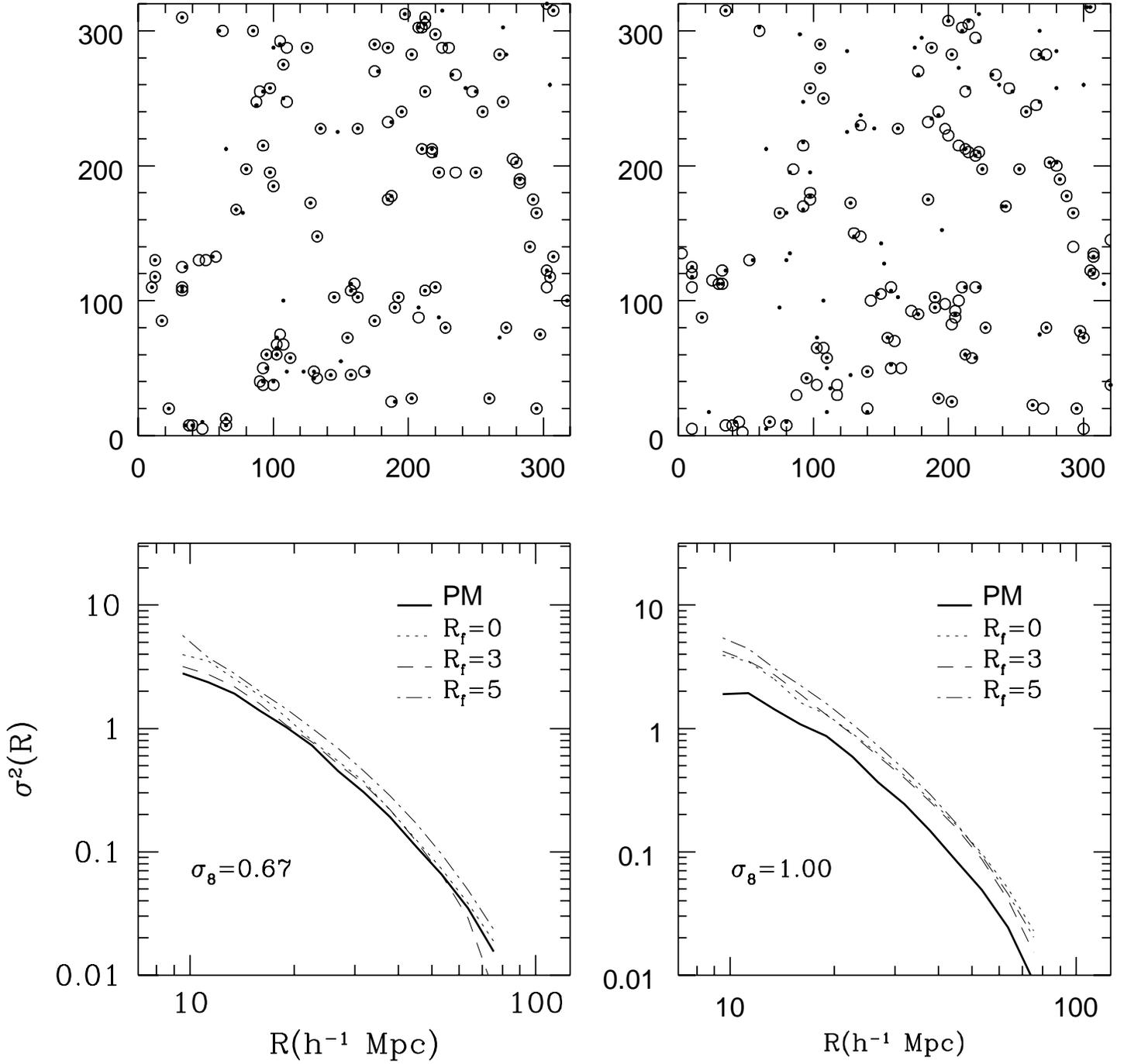

# The Cluster Distribution as a Test of Dark Matter Models. I: Clustering Properties

Stefano Borgani[1,2], Manolis Plionis[2], Peter Coles[3] and Lauro Moscardini[4]
[1] *INFN – Sezione di Perugia, c/o Dipartimento di Fisica dell'Università, via A. Pascoli, I-06100 Perugia, Italy*
[2] *SISSA – International School for Advanced Studies, via Beirut 2-4, I-34013 Trieste, Italy*
[3] *Astronomy Unit, School of Mathematical Sciences, Queen Mary & Westfield College, Mile End Road, London E1 4NS, UK*
[4] *Dipartimento di Astronomia, Università di Padova, vicolo dell'Osservatorio 5, I-35122 Padova, Italy*

10 May 1995

**ABSTRACT**
We present extended simulations of the large–scale distribution of galaxy clusters in several dark matter models, using an optimized version of the truncated Zel'dovich approximation (TZA). In order to test the reliability of our simulations, we compare them with N–body based cluster simulations. We find that the TZA provides a very accurate description of the cluster distribution as long as fluctuations on the cluster mass scale are in the mildly non–linear regime ($\sigma_8 \lesssim 1$). The low computational cost of this simulation technique allows us to run a large ensemble of fifty realizations for each model, so we are able to quantify accurately the effects of cosmic variance. Six different dark matter models are studied in this work: Standard CDM (SCDM), Tilted CDM (TCDM) with primordial spectral index $n = 0.7$, Cold + Hot DM (CHDM) with $\Omega_{hot} = 0.3$, low Hubble constant ($h = 0.3$) CDM (LOWH) and two spatially flat low–density CDM models with $\Omega_\circ = 0.2$ and $\Omega_\Lambda = 0.8$, having two different normalizations, $\sigma_8 = 0.8$ ($\Lambda\mathrm{CDM}_1$) and $\sigma_8 = 1.3$ ($\Lambda\mathrm{CDM}_2$). We compare the cluster simulations with an extended redshift sample of Abell/ACO clusters, using various statistical measures, such as the integral of the two–point correlation function, $J_3$, and the probability density function (pdf). We find that the models that best reproduce the clustering of the Abell/ACO cluster sample are the CHDM and the $\Lambda\mathrm{CDM}_1$ models. The $\Lambda\mathrm{CDM}_2$ model is too strongly clustered and this is probably overestimated in our simulations due to the large $\sigma_8$ value of this model. All the other models are ruled out at a high confidence level. The pdfs of all models are well approximated by a lognormal distribution, consistent with similar findings for Abell/ACO clusters. The low–order moments of all the pdfs are found to obey a variance–skewness relation of the form $\gamma \approx S_3 \sigma^4$, with $S_3 \simeq 1.9$, independent of the primordial spectrum shape and consistent with observational data. After computing the cluster biasing parameter, $b_{cl}$, we estimate the quantity $\beta_{cl} = \Omega_\circ^{0.6}/b_{cl}$ for the different models. Owing to the rather large observational uncertainties, $\beta_{cl} = 0.20 \pm 0.05$, this test does not discriminate strongly between the different models. The scale–independence of $\beta_{cl}$, and thus of $b_{cl}$, does, however, suggest that it is probably a reliable procedure to use the linear biasing model to infer the dark matter power–spectrum from observational cluster samples. We also note that the abundances of clusters predicted using the Press–Schechter theory provide strong constraints on these models: only the CHDM, LOWH and $\Lambda\mathrm{CDM}_2$ models appear to produce the correct number–density of clusters.

## 1 INTRODUCTION

The study of the distribution of matter on the largest scales provides important constraints on models of cosmic structure formation. If the gravitational instability picture is correct, the expected displacements of clusters of galaxies away from their primordial positions are much smaller than the typical separation of these objects. In principle, therefore, clusters of galaxies can yield clues about the primordial spectrum of perturbations that gave rise to them. This is the reason why so much effort has been devoted to compiling deep cluster surveys, starting with the pioneering work of Abell (1958), Zwicky et al. (1968) and Abell, Corwin & Olowin (1989), and leading up to extended redshift surveys both in the optical (e.g. Postman, Huchra & Geller 1992; Dalton et al. 1994; Collins et al. 1994, and references therein) and in the X–ray (e.g. Nichol, Briel & Henry 1994; Romer et al. 1994) regions of the spectrum.

Accompanying the observational challenge of acquiring extended cluster redshift surveys, a great deal of effort has also been directed towards the provision of reliable statisti-cal characterizations of the cluster distribution. It has been established that the cluster two–point correlation function is well modelled by a power–law,

$$\xi(r) \; = \; (r/r_o)^{-\gamma}. \tag{1}$$

Although the slope, $\gamma \simeq 1.8$, turns out to be quite similar to that of galaxies, the correlation length, $r_o$, is much larger (cf. Bahcall & Soneira 1983; Klypin & Kopylov 1983; Bahcall 1988). Different determinations, based on different cluster samples, indicate values in the range $r_0 = 13-25\,h^{-1}Mpc^{\star}$ (e.g., Nichol et al. 1994, and references therein), while more recently the reliability of the power–law model for $\xi(r)$ has also been questioned by different authors (e.g. Olivier et al. 1993).

In order to compare the observational cluster data sets with different cosmological models several authors have resorted to large N–body simulations which were designed to sample the length–scales relevant to the cluster distribution (e.g. White et al. 1987; Bahcall & Cen 1992; Croft & Efstathiou 1994). The problem with this kind of approach is that large N–body simulations are very expensive from a computational point of view. Therefore, one is usually forced to consider only a limited number of models, with a small number of independent realizations for each model. In this respect, analytical approaches, based either on Eulerian linear theory (e.g. Bardeen et al. 1986; Coles 1989; Lumsden, Heavens & Peacock 1989; Borgani 1990; Holtzman & Primack 1993) or on the Zel'dovich approximation (ZA, hereafter; e.g. Doroshkevich & Shandarin 1978; Mann, Heavens & Peacock 1993), are in general preferred to numerical simulations. Nevertheless, they are of limited utility since one is often obliged to resort to oversimplifying assumptions about the nature of galaxy clusters. Furthermore, statistics which go beyond the two–point correlation function and its Fourier transform, the power–spectrum, are hard to handle. Finally, it is not clear how shot–noise effects and/or observational biases (e.g., redshift–space distortions, selection functions, non–trivial sample geometry) can be realistically modelled in order to allow a consistent comparison with real data sets.

In a previous paper (Borgani, Coles & Moscardini 1994, hereafter Paper I) we used the truncated Zel'dovich approximation (TZA hereafter) to generate cluster simulations which were accurate when compared to N–body simulations, and, at the same time, computationally so cheap as to enable us to produce many realizations of several dark matter (DM) models (see also Blumenthal, Dekel & Primack 1988). Recently, Sathyaprakash et al. (1994) compared the TZA and several approximations of non–linear gravitational clustering to direct N–body results. They showed that, although the TZA fails to follow small–scale clustering in the multistream region, it is nevertheless able to account for non–local effects due to long wavelength modes in the density fluctuation spectrum. For this reason, the TZA is not expected to provide a correct description of the internal structure and mass distribution of non–linear structures like galaxy clusters, but it should be accurate in locating their correct positions and thus reliably describes their spatial distribution. In this respect, it is not necessary to employ the full power and sophistication of modern N–body methods to investigate cluster clustering on large scales. It should be clear that cluster positions will be accurately determined as long as fluctuations on the cluster mass scale are in the linear or mildly non–linear regime ($\sigma_8 \lesssim 1$). However, at later epochs, when $\sigma_8 > 1$, shell crossing is no longer negligible on such scales and non–linear effects such as cluster infall and merging become relevant.

Given the approximate nature of our method, it is particularly important to test its reliability for the task we have set it. Indeed, in a recent paper, Gaztañaga, Croft & Dalton (1995) have compared cluster simulations based on the TZA and on N–body for a model with $\sigma_8 = 1.3$. Based on their analysis of this one model, they reached the general conclusion that TZA cluster simulations are not reliable for simulating the cluster distribution under any circumstances. One of the purposes of the present paper is to demonstrate that the objections of Gaztañaga et al. apply only to models in which $\sigma_8$ is particularly large and that our technique is actually extremely accurate for models in which $\sigma_8$ is of order unity or less. To do this, we present cluster simulations based on an implementation of the ZA that has been substantially improved with respect to that used in Paper I (see also Plionis et al. 1995, hereafter Paper II). Indeed, we can even show that our method is as accurate as $N$–body methods in situations where its use is appropriate and has a fraction of the computational cost. This low cost has allowed us to generate 50 realizations of each of six different initial power–spectra for statistical analysis. This quantitative analysis is the second purpose of this paper.

The availability of such a large set of simulations represents an extremely powerful test–bed for constraining DM models through a detailed comparison of the statistical properties of real and simulated cluster distributions. However, a reliable quantitative measure of cluster clustering is not easy to find. Clusters are rather rare objects with typical mean separation of several tens of Mpcs; while bright galaxies have a mean separation comparable to their correlation length, clusters have a mean separation which is twice the corresponding $r_o$ value. For this reason, shot–noise effects become important on small scales ($\lesssim 10\,h^{-1}Mpc$) while, on larger scales ($\gtrsim 40\,h^{-1}Mpc$), a low signal–to–noise ratio is expected because the clustering is weak. Robust statistical estimators, which are able to provide reliable measures over a large range of scales, are required to describe the cluster distribution properly and to allow an effective comparison with model predictions.

As a first statistical test, we will use in this paper the quantity $J_3(R)$, which is defined through the integral of $\xi(r)$:

$$J_3(R) \; = \; \frac{1}{4\pi} \int_0^R \xi(r)\, r^2\, dr. \tag{2}$$

The advantage of using $J_3(R)$ over $\xi(r)$ lies in the fact that in a sparse distribution of objects, an integral quantity such as that defined by eq. (2), should be less susceptible to statistical noise than a differential one, such as $\xi(r)$.

An alternative method, which is becoming increasingly popular, is the study of the probability density function (pdf) itself. Usually one attempts to obtain a continuous density field by smoothing the discrete distribution of objects with some window function (a top–hat or a Gaus-sian are the most commonly used). The smoothing procedure itself significantly reduces the shot–noise, which could dominate the discrete distribution (Gaztañaga & Yokoyama 1993). Then one can define the pdf $f(\varrho)$, where $\varrho = \rho/\langle\rho\rangle$, and derive its moments, defined by:

$$\langle\delta^n\rangle = \int_{-1}^{\infty} \delta^n f(\delta) d\delta, \qquad (3)$$

where $\delta = \varrho - 1$.

---

$^{\star}$ $h$ is the Hubble constant in units of $100\,\mathrm{km\,s^{-1}\,Mpc^{-1}}$.

The pdf and moments of different galaxy samples have been estimated by various authors (e.g. Saunders et al. 1991; Bouchet et al. 1993; Gaztañaga and Yokoyama 1993; Sheth, Mo & Saslaw 1994). Kofman et al. (1994) have compared the pdf derived from CDM N–body simulations with that of the IRAS sample and the one recovered using the POTENT procedure with $\Omega = 1$ (see also Lahav et al. 1993). Their main conclusion is that, if galaxies trace the mass, the observed pdf is consistent with Gaussian initial conditions.

Plionis & Valdarnini (1995, hereafter PV95) studied the pdf (and its moments) of the 3–D Abell/ACO smoothed cluster distribution and compared them with static simulations, based on a Gaussian fluctuation spectrum, which reproduced the two– and three–point cluster correlation functions as well as the observed selection effects. They found that the real and simulated cluster pdfs are well approximated by a lognormal distribution. Cappi & Maurogordato (1995) have realized a study of the higher–order moments for the discrete Abell/ACO cluster distribution, for both projected and redshift samples, while Kolatt, Dekel & Primack (1995) estimated the pdf for real cluster samples as well as for cluster N–body simulations based both on Gaussian and non–Gaussian initial CDM fluctuations. They concluded that no evidence of non–Gaussian initial conditions are imprinted into the shape of the cluster pdf. In our Paper II we compared the variance and the skewness of the smoothed Abell/ACO cluster pdf with those obtained from the TZA simulations for a list of DM models. In this paper we will study the pdf statistics of our latest cluster simulations. Using the same analysis procedure as that used for the Abell/ACO cluster sample (PV95), we will put stringent constraints on the models we consider.

The layout of this paper is as follows. In Section 2 we describe the cluster simulations, i.e., how to optimize the ZA, the method of cluster identification, the simulation reliability when compared to N–body and higher–order Lagrangian results, and the considered models for the power–spectrum. In view of the unsuitability of our simulation method for studying the small–scale structure of clusters, we also use the Press & Schechter (1974; PS hereafter) method to compare the cluster abundances predicted by DM models with the available observational data. This analysis provides an independent constraint on the models we are considering. In Section 3 we describe the Abell/ACO sample we use. In Section 4 we present the analysis of the discrete cluster distribution using the $J_3$ integral, while in Section 5 we present the pdf and moment analysis of the smoothed cluster distribution. In Section 6 we discuss our results and state our main conclusions.

## 2 THE SIMULATIONS

### 2.1 The Zel'dovich approach

The Zel'dovich approximation (Zel'dovich 1970; Shandarin & Zel'dovich 1989) is based on the assumption of laminar flow for the motion of a self–gravitating non–relativistic collisionless fluid. Let $\mathbf{q}$ be the initial (Eulerian) position of a fluid element and $\mathbf{r}(\mathbf{q}, t) = a(t)\mathbf{x}(\mathbf{q}, t)$ the final position at the time $t$, which is related to the comoving Lagrangian coordinate $\mathbf{x}(\mathbf{q}, t)$ through the cosmic expansion factor $a(t)$. The ZA amounts to assume the expression

$$\mathbf{r}(\mathbf{q}, t) = a(t) \left[\mathbf{q} + b(t)\nabla_\mathbf{q}\psi(\mathbf{q})\right] \qquad (4)$$

for the Eulerian–to–Lagrangian coordinate mapping. In eq.(4) $b(t)$ is the growing mode for the evolution of linear density perturbations and $\psi(\mathbf{q})$ is the gravitational potential, which is related to the initial density fluctuation field, $\delta(\mathbf{q})$, through the Poisson equation

$$\nabla^2\psi(\mathbf{q}) = -\frac{\delta(\mathbf{q})}{a(t)}. \qquad (5)$$

As a result of the factorization of the $t$– and $\mathbf{q}$– dependence in the displacement term of eq.(4), the fluid particles move under this approximation along straight lines, with comoving peculiar velocity

$$\mathbf{v}(\mathbf{q}, t) = \dot{\mathbf{x}}(\mathbf{q}, t) = \dot{b}(t)\nabla_\mathbf{q}\psi(\mathbf{q}). \qquad (6)$$

Therefore, gravity determines the initial kick to the fluid particles through eqs.(5) and (6), and afterwards they do not feel any tidal interactions. Particles fall inside gravitational wells to form structures, which however quickly evaporate. In this sense, the ZA gives a good description of gravitational dynamics as far as particle trajectories do not intersect with each other, while its validity breaks down when shell–crossing occurs, and local gravity dominates.

Several prescriptions have been suggested to overcome the shortcomings of the ZA, such as adding a small viscous term to the equation of motion for the fluid, or by going to higher–orders in Lagrangian perturbative theory (e.g. Sahni & Coles 1995, and references therein). As a further possibility, Coles, Melott & Shandarin (1993) have shown that filtering out the small–scale wavelength modes in the linear power–spectrum reduces the amount of shell–crossing, thus improving the performance of the ZA. Melott, Pellman & Shandarin (1993) claimed that an optimal filtering procedure is obtained by convolving the linear power–spectrum with the Gaussian filter

$$W_G(kR_f) = e^{-(kR_f)^2/2}. \qquad (7)$$

The problem then arises of choosing the filtering radius $R_f$ appropriately, in order to suppress shell–crossing as much as possible without preventing genuine clustering to build up. In Paper I we chose $R_f$ so that the expected mass within a Gaussian window of that radius were of the same order ($\sim 10^{15} M_\odot$) of the mass for a rich galaxy cluster. The disadvantage of this approach is that it does not rely on any objective criterion to optimize the ZA and treats in the same fashion different fluctuation spectra, which should produce a different amount of shell–crossing. The resulting filtering radius, $R_f = 5\Omega_0^{-1/3} h^{-1} Mpc$, is generally larger than the optimal ones, which we use in the present paper, thus causing an excessive removal of clustering.

Kofman et al. (1994) derived an analytical expression for the average number of streams at each Eulerian point, $N_s$, as a function of the r.m.s. fluctuation level of the initial Gaussian density field. In Figure 1 we plot $N_s$ as a function of the filtering scale $R_f$ for the six different power-spectra that we will consider (see next subsection), evaluated according to eq. (7) of Kofman et al. (1994). As a general criterion, we decided to choose $R_f$ for each model so that $N_s = 1.1$. We found this to be a reasonable compromise between smaller $N_s$ values, giving rapidly increasing $R_f$ and high suppression of clustering, and larger $N_s$, at which the ZA progressively breaks down. The resulting r.m.s. fluctuation value corresponding to $N_s = 1.1$ is $\sigma = 0.88$.

By adopting this implementation of the TZA, the main steps of our cluster simulations are the following:

(a) Convolve the linear power-spectrum with the Gaussian window of eq. (7) and $R_f$ chosen as previously described.
(b) Generate a random-phase realization of the density field on $128^3$ grid points for a cubic box of $L = 320 \, h^{-1} Mpc$ aside.
(c) Move $128^3$ particles having initial Lagrangian position on the grid, according to the TZA. Each particle carries a mass of $4.4 \times 10^{12} h^{-1} \Omega_\circ M_\odot$.
(d) Reassign the density and the velocity field on the grid through a TSC interpolation scheme (e.g. Hockney & Eastwood 1981) for the mass and the moment carried by each particle.
(e) Select clusters as local density maxima on the grid according to the following prescription. If $d_{cl}$ is the average cluster separation, then we select $N_{cl} = (L/d_{cl})^3$ clusters as the $N_{cl}$ highest density peaks. In the following, we assume $d_{cl} = 40 \, h^{-1} Mpc$, which is appropriate for the combined Abell/ACO cluster sample to which we will compare our simulation results (see Section 3). Therefore, we will analyze a distribution of 512 clusters in each simulation box, with periodic boundary conditions.

## 2.2  Reliability of the TZA

Before entering into the presentation of our analysis, we present the results of detailed tests of the reliability of TZA for simulating the large-scale distribution of galaxy clusters. For this test, we ran a PM simulation for an initial spectrum corresponding to the Cold+Hot DM (CHDM) model with $\Omega_{hot} = 0.3$ (see Section 2.3 for more details). The size of the simulation box and the number of grid points and particles were identical to those for the TZA simulations. We do not attempt here to distinguish between hot and cold particles, since the simulation is intended only for a comparison with the Zel'dovich approach. In any case, it is reasonable to assume that the adopted mass resolution is low enough for effects of neutrino free-streaming to be negligible. Clusters are identified as local maxima on the grid, following the same prescription outlined in Section 2.1. We compare the outputs of the N-body simulation to those of several TZA realizations, each based on the same initial phase assignment, but having different filtering radii $R_f$.

In Figure 2 we compare the cluster distributions within a slice $80 \, h^{-1} Mpc$ thick and the cluster count-in-cell variance, $\sigma^2(R)$, for PM and TZA simulations, at two different evolutionary stages, corresponding to $\sigma_8 = 0.67$ and $\sigma_8 = 1$ for the r.m.s. fluctuation amplitude within a top-hat sphere of $8 \, h^{-1} Mpc$ radius. The less-evolved stage is consistent with the two-year COBE normalization supplied by Bennett et al. (1994), but the later stage corresponds to a higher normalization than this. We consider the second stage only in order to assess the reliability of the TZA simulations when non-linear effects appear on the cluster mass scale.

The variance has been estimated according to

$$\sigma^2(R) = \frac{\langle N^2 \rangle_R - \langle N \rangle_R^2}{\langle N \rangle_R^2} - \frac{1}{\langle N \rangle_R}, \quad (8)$$

where $\langle N \rangle_R$ and $\langle N^2 \rangle_R$ are the average and the second order moment, respectively, for counts within 20,000 randomly placed spheres of radius $R$ and the second term in the l.h.s. of eq.(8) represents the correction for Poissonian shot-noise. We take $R$ between $\sim 10$ and $80 \, h^{-1} Mpc$, since smaller scales are heavily affected by shot-noise and larger scales by effects of periodic boundary conditions.

At $\sigma_8 = 0.67$ the TZA reproduces the variance of the PM cluster distribution remarkably well, especially when the linear power-spectrum is mildly filtered with $R_f = 3 \, h^{-1} Mpc$, a value which is quite close to that required for $N_s = 1.1$. For larger filtering no allowance is made for genuine clustering to build up. As a consequence, the large-to-small scale power ratio is decreased, with a subsequent increase of the cluster correlation. On the other hand, a similar clustering suppression is caused by shell-crossing when a smaller filtering, or no filtering at all, is taken. It should be stressed, however, that the agreement between N-body and TZA cluster simulations is not merely in a statistical sense. In the upper panel we superimpose the cluster distributions from PM (filled dots) and TZA ($R_f = 3 \, h^{-1} Mpc$; open circles) simulations. The relative positions are identical for almost all the clusters. This confirms that the TZA places peaks in the correct positions when it is suitably "truncated", thus disproving the general conclusion by Gaztañaga et al. (1995) that the TZA is not reliable enough for cluster simulations.

In order to verify that the close agreement between TZA and PM cluster simulations is generally attained, even when considering N-body simulations with different dynamical and mass resolution, we compare in Figure 3 the two-point cluster correlation function for our CHDM simulations (open dots) to that obtained by Klypin & Rhee 1994; KR94, hereafter) by evolving until $\sigma_8 = 0.67$ the same spectrum with a PM N-body code with $256^3$ grid points and particles within a box of $200 \, h^{-1} Mpc$. The KR94 results are obtained as an average over 2 realizations and error bars are quasi-Poissonian sampling uncertainties while our results refer to the average taken over 50 random realizations. Our error bars are estimated as the r.m.s. scatter over this ensemble. By comparing this plot with Fig. 2 of Paper I, it is apparent that we have improved our implementation of the TZA by increasing the resolution and by optimizing the power-spectrum filtering. The agreement between the TZA and KR94 results is really remarkable and further confirms the reliability of the TZA to follow correctly the mildly non-linear clustering regime.

This agreement, however, is not so good when later evolutionary stages are considered. Indeed, at $\sigma_8 = 1$, although $\sigma^2(R)$ for TZA clusters increases to some extent, it is rather stable or marginally decreasing for PM clusters. As a result, the TZA systematically overestimates the cluster correlations for any choice of the filtering radius. The lack of evolution of $\sigma^2(R)$ in the N-body simulations is more likely to be ascribed to a balance between two competitive effects, the infall of close clusters, which tends to enhance the small-scale clustering, and cluster merging, which decreases the number of close pairs, rather than to a freezing of the clustering. This kind of picture is also supported by studies of the correlation function of cluster peculiar velocities (Cen, Bahcall & Gramann 1994; Croft & Efstathiou 1995), which has negative values below $20\,h^{-1}\,Mpc$, showing therefore evidence of infall on such scales. In the N-body treatment and in terms of the cluster distribution, the disappearance of close pairs within overdense regions leads one to pick up relatively more clusters in the "field", as one is willing to select a fixed number of such clusters (cf. the upper right panel in Figure 2).

We conclude from this analysis that the TZA is very accurate for cluster simulations as long as effects of non-linear gravitational dynamics are negligible on the cluster mass scale (i.e., $\sigma_8 \lesssim 1$). Although this holds in most cases of cosmological relevance (for instance, $\sigma_8 \simeq 0.6$ is required for $\Omega_\circ = 1$ CDM like models to fit the observed cluster abundance; cf. White, Efstathiou & Frenk 1993, and Section 2.4 below), there are nevertheless a few exceptions, for which the conditions for our method to be accurate are not satisfied: we should therefore bear in mind the above words of caution.

On the basis of this result, one may wonder whether including gravitational effects in the ZA, at least to first order, would improve the agreement with N-body for $\sigma_8 = 1$. To this purpose, we realised simulations based on the second-order Lagrangian theory (e.g., Moutarde et al. 1991; Bouchet et al. 1992; Buchert & Ehlers 1993; Gramann 1993; Catelan 1995). Expanding equation (4), for the particle displacements, to the second order in $b(t)$ we obtain

$$\mathbf{r}(\mathbf{q},t) = a(t)\left[\mathbf{q} + b(t)p(\mathbf{q}) + b^2(t)s(\mathbf{q})\right]. \quad (9)$$

Here, $p(\mathbf{q}) = \nabla_\mathbf{q} \psi(\mathbf{q})$ is the first order term of eq.(4), while, for $\Omega = 1$, the term

$$s(\mathbf{q}) = -\frac{3}{7} \sum_{1 \le i < j \le 3} \left(\frac{\partial p_i}{\partial q_j}\frac{\partial p_j}{\partial q_i} - \frac{\partial p_i}{\partial q_i}\frac{\partial p_j}{\partial q_j}\right) \quad (10)$$

includes tidal effects, which generate the first-order deviation from the straight-line particle trajectories.

In Figure 4 we plot the cluster two-point correlation function, $\xi(r)$, for the CHDM simulations, based on the first-order (filled circles) and on the second-order (open circles) TZA, for both $\sigma_8 = 0.67$ (left panel) and $\sigma_8 = 1$ (right panel). Results correspond to the average over 10 realizations and error bars are $1\sigma$ scatter over this ensemble. Note that going to second-order in Lagrangian theory does not introduce any significant change in the cluster distribution. This suggests that non-linear gravitational effects, like merging and infall of structures, cannot be treated with a perturbative approach, at least at this order.

### 2.3 Dark matter models

We ran simulations for six different models of the initial fluctuation spectrum. For each model, we generate 50 random realizations, so as to reliably estimate the effect of cosmic variance. All the models, except the $\Lambda \text{CDM}_1$ one (see below), are normalized to be consistent with the COBE measured quadrupole of CMB temperature anisotropy (Bennett et al. 1994). The models we have considered are the following.

(1) The standard CDM model (SCDM), with $\Omega_\circ = 1$, $h = 0.5$ and $\sigma_8 = 1$ for the r.m.s. fluctuation amplitude within a top-hat sphere of $8\,h^{-1}\,Mpc$. The rather large normalization of this model could produce an overestimate of cluster clustering in the light of the previous discussion about the reliability of the TZA cluster simulations. However, as we shall see, this model already produces too weak clustering for clusters, while a more accurate treatment would eventually produce even weaker correlations. Therefore we will give at most only an underestimate of the confidence level to which SCDM should be ruled out.

(2) A tilted CDM model (TCDM), with $n = 0.7$ for the primordial spectral index. Tilting the primordial spectral shape from the scale-free one has been suggested in order to improve the CDM description of the large-scale structure (e.g. Cen et al. 1992; Tormen et al. 1993; Liddle & Lyth 1993; Adams et al. 1993; Moscardini et al. 1995).

(3) A low Hubble constant CDM model (LOWH), with $h = 0.3$. Decreasing the Hubble constant has the effect of increasing the horizon size at the equivalence epoch, thus pushing the turnover of the spectrum to its scale-free form out to larger scales. The relevance of this model in alleviating several cosmological problems has been recently emphasized by Bartlett et al. (1994).

(4) A Cold + Hot DM model (CHDM), with $\Omega_{hot} = 0.3$ for the fractional density contributed by the hot particles. For a fixed large-scale normalization, adding a hot component has the effect of suppressing the power-spectrum amplitude at small wavelengths (see, e.g., Klypin et al. 1993; Klypin, Nolthenius & Primack 1995, and references therein, for the relevance of CHDM). Although the small-scale peculiar velocities are lowered to an adequate level, the corresponding galaxy formation time is delayed so that such a model is strongly constrained by the detection of high-redshift objects (e.g. Ma & Bertschinger 1994; Klypin et al. 1995, and references therein).

(5) A spatially flat, low-density CDM model ($\Lambda \text{CDM}_1$), with $\Omega_\circ = 0.2$, $\Omega_\Lambda = 0.8$ for the cosmological constant term (e.g., Bahcall & Cen 1992; Baugh & Efstathiou 1993; Peacock & Dodds 1994) and $\sigma_8 = 0.8$. With this normalization, $\Lambda \text{CDM}_1$ has a significantly lower amplitude than that implied by COBE data.

(6) The same model as in (5), but with a larger normalization, $\sigma_8 = 1.3$ ($\Lambda \text{CDM}_2$), so as to be consistent with COBE results. Having in mind the comparison between PM and TZA simulations previously discussed, the $\Lambda \text{CDM}_2$ model is expected to have a too large $\sigma_8$ value to be adequately treated by the TZA. For this reason, we prefer to consider also the $\Lambda \text{CDM}_1$ model, whose lower normalization allows it to be properly handled by the TZA. Since cluster correlations have been shown to be almost independent of $\sigma_8$, we could argue that results at $\sigma_8 = 0.8$ can be considered as representative of those at $\sigma_8 = 1.3$.

**Table 1.** The models. Column 2: the density parameter $\Omega_0$; Column 3: the cosmological constant term $\Omega_\Lambda$; Column 4: the density parameter of the hot component $\Omega_{hot}$; Column 5: the primordial spectral index $n$; Column 6: the Hubble parameter $h$; Column 7: the linear r.m.s. fluctuation amplitude at $8\,h^{-1}Mpc$ $\sigma_8$; Column 8: The filtering radius, $R_f$, in units of $h^{-1}Mpc$, corresponding to $N_s = 1.1$ for the level of orbit crossing.

| Model | $\Omega_0$ | $\Omega_\Lambda$ | $\Omega_{hot}$ | $n$ | $h$ | $\sigma_8$ | $R_f$ |
|---|---|---|---|---|---|---|---|
| SCDM | 1.0 | 0.0 | 0.0 | 1.0 | 0.5 | 1.0 | 4.4 |
| TCDM | 1.0 | 0.0 | 0.0 | 0.7 | 0.5 | 0.5 | 1.6 |
| LOWH | 1.0 | 0.0 | 0.0 | 1.0 | 0.3 | 0.6 | 2.4 |
| CHDM | 1.0 | 0.0 | 0.3 | 1.0 | 0.5 | 0.7 | 2.2 |
| $\Lambda$CDM$_1$ | 0.2 | 0.8 | 0.0 | 1.0 | 1.0 | 0.8 | 3.2 |
| $\Lambda$CDM$_2$ | 0.2 | 0.8 | 0.0 | 1.0 | 1.0 | 1.3 | 6.3 |

The transfer functions for the above models have been taken from Holtzman (1989), except that of LOWH, which is taken from Bond & Efstathiou (1984), with suitably chosen shape parameter $\Gamma = \Omega_o h = 0.3$. We note that the latter transfer function assumes the baryonic component to be negligible. However, for $h = 0.3$, constraints from the Big Bang nucleosynthesis imply a considerable baryon fraction, $\Omega_b \simeq 0.15$ (e.g., Walker et al. 1991). Although this could be a limitation for small scale considerations, it hardly affect the results at the large scales we are interested in. All the model parameters are listed in Table 1.

In Paper II we also considered a purely open CDM model with $\Omega_o = 0.2$, without cosmological constant and with $\sigma_8 = 1$. However the only difference between low–density models with and without cosmological constant lies in the different way one normalizes to the large–scales CMB anisotropies, while the spectrum shapes are virtually identical. Therefore, we expect this model to be just intermediate between the two above $\Lambda$CDM models.

Each power–spectrum is suitably smoothed on the scale $R_f$ according to the prescription described in Section 2.1. In Figure 1 we plot the average stream number per Eulerian point, $N_s$, as a function of the filtering radius for all the above models. The intersection of the $N_s = 1.1$ line with each curve indicates the smoothing scale adopted for the corresponding spectrum (see also Table 1). Note that the larger $\sigma_8$ value for $\Lambda$CDM$_2$ requires a stronger filtering to suppress shell–crossing.

In Figure 5 we plot the projected particle distribution within a slice $10\,h^{-1}Mpc$ thick, superimposing the cluster distribution, for the CHDM model. In order to better show how the identified clusters trace the underlying density field, we used simulations within a $640\,h^{-1}Mpc$ box. As expected, clusters are preferentially located at the knots corresponding to the intersections between filaments, while they avoid long filaments and flattened pancakes, so as to generate devoid regions of size $\sim 200\,h^{-1}Mpc$, at least within the slice.

### 2.4 Cluster Abundances

As an independent constraint on the above models, we have computed the expected cluster abundances, as predicted by the standard PS formalism. If structures are identified through a filter $W$ on a scale $R$, in order to have a mass $M = f\bar{\rho}R^3$ ($\bar{\rho}$ is the average matter density), the PS formula for the number density of objects with mass between $M$ and $M + dM$ is

$$n(M)\,dM = \frac{\delta_c}{f\sqrt{2\pi}} \int_R^\infty \frac{\eta(R)}{\sigma(R)} \exp\left(-\frac{\delta_c^2}{2\sigma^2(R)}\right) \frac{dR}{R^2}, \quad (11)$$

where

$$\eta(R) = \frac{1}{2\pi^2 \sigma^2(R)} \int k^4\, P(k)\, \frac{dW^2(kR)}{d(kR)}\, \frac{dk}{kR}\, ;$$

$$\sigma^2(R) = \frac{1}{2\pi^2} \int k^2\, P(k)\, W^2(kR)\, dk\,. \quad (12)$$

Therefore, the total abundance of objects of mass larger than $M$ is

$$N(>M) = \int_M^\infty n(M')\,dM'\,. \quad (13)$$

In the above expressions $f$ is a "form factor", which depends on the shape of the filter $W$: $f = (2\pi)^{3/2}$ for a Gaussian filter, and $f = 4\pi/3$ for a top-hat filter. A Gaussian filter will be assumed in the following analysis. The parameter $\delta_c$ is the critical density contrast, which represents the threshold value for a fluctuation to turn into an observable object, if evolved to the present time by linear theory. Arguments based on a simple spherical collapse suggest $\delta_c = 1.68$, but the inclusion of non–linear effects, as well as aspherical collapse, may lead to a lower value of $\delta_c$. For example, Klypin & Rhee (1994; KR94 hereafter) found that the cluster mass function in their CHDM N–body simulations is well fitted by eq.(11) by taking $\delta_c = 1.5$.

White, Efstathiou & Frenk (1993) resorted to X–ray data for the temperatures of the gas component of clusters and estimated a cluster abundance of about $4 \times 10^{-6}(h^{-1}Mpc)^{-3}$ for masses exceeding $M = 4.2 \times 10^{14} h^{-1} M_\odot$. Using observed cluster velocity dispersion, Biviano et al. (1993) obtained an abundance of about $6 \times 10^{-6}(h^{-1}Mpc)^{-3}$ for clusters exceeding the above mass limit.

In Figure 6 we compare model predictions at different $\delta_c$ values to the above observational estimates. Note that realistic uncertainties on cluster abundances are probably larger than the difference between the two above values. They should include variations in the average cluster number density between different samples, biases toward high mass for observations of cluster velocity dispersion, uncertainties in the model used to relate gas temperature and cluster mass, etc. Keeping in mind such warnings, we note from Figure 6 that $\Lambda$CDM$_1$ is ruled out; for $\delta_c = 1.5$ it produces more than one order of magnitude less clusters than observed. This agrees with the suggestion of White et al. (1993), that a higher normalization ($\sigma_8 \simeq 1.4$) is required for such a low–density model to provide a correct number of clusters. Consistently, $\Lambda$CDM$_2$ gives the right cluster abundance for reasonable values of $\delta_c$. As for the $\Omega_o = 1$ models, it turns out that the resulting abundances depend mostly on the $\sigma_8$ normalization value and not on the shape of the spectrum. This is not surprising: the Gaussian smoothing scale, $R \simeq 4.6\, h^{-1}\, Mpc$, which encompasses a mass of $4.2 \times 10^{14}\, h^{-1}\, M_\odot$, is equivalent to a top-hat sphere of about $7\, h^{-1}\, Mpc$, which is rather close to the normalization scale. As a result, SCDM turns out to produce too many clusters for any reasonable value of $\delta_c$. On the other hand, the low normalization of TCDM turns into a severe underproduction of clusters, even at the smallest $\delta_c$ values. The only two flat models which generate cluster abundances in agreement with the results by White et al. (1993) and Biviano et al. (1993) for a reasonable choice of $\delta_c$ are LOWH and CHDM.

A note of caution about the strength of the constraints emerging from the Press–Schechter analysis, is due for a number of reasons. First, there is some evidence of a discrepancy between the mass profiles of clusters inferred from X-ray data (e.g. Edge & Stewart 1991) and from gravitational lensing considerations (e.g. Kaiser et al. 1994), so it is not clear whether the cluster mass function inferred from the X-ray data is correct. Likewise, there is a possibility that the distribution of cluster peculiar velocities may be affected by subclustering. One should also mention that the applicability of the Press–Schechter method is itself open to some doubt. Although, as we mentioned above, it appears to perform well for the CHDM model when compared with N-body simulations, its accuracy is yet to be verified for the other models. In the case of LOWH, where a much higher fraction of the cluster mass is baryonic than in the other models, one might imagine this formalism to be particularly suspect. We therefore take the constraints emerging from this analysis to be indicative but not watertight.

## 3 THE CLUSTER SAMPLE

We use the combined Abell/ACO $R \geq 0$ cluster sample, as defined in Plionis & Valdarnini (1991) [hereafter PV91] and analysed in Plionis, Valdarnini & Jing (1992) [hereafter PVJ] and Plionis & Valdarnini (1995). The northern sample, with dec $\geq -17°$ (Abell), is defined by those clusters that have measured redshift $z \lesssim 0.1$, while the southern sample (ACO; Abell, Corwin & Olowin 1989), with dec$\leq -17°$, is defined by those clusters with $m_{10} \leq 16.4$ (note that with this definition and due to the availability of many new cluster redshifts only 7 ACO clusters have $m_{10}$ estimated redshifts from the $m_{10}-z$ relation derived in PV91). Both samples are limited in Galactic latitude by $|b| \geq 30°$. The redshifts have been taken from a number of studies, the references of which can be found in PV95. The total number of clusters in our samples is 357 and 157, for Abell and ACO respectively.

To take into account the effect of Galactic absorption, we assume the usual cosecant law:

$$P(|b|) = \text{dex}\,[\alpha\,(1 - \csc|b|)] \qquad (14)$$

with $\alpha \approx 0.3$ for the Abell sample (Bahcall & Soneira 1983; Postman et al. 1989) and $\alpha \approx 0.2$ for the ACO sample (Batuski et al. 1989). The cluster–redshift selection function, $P(z)$, is determined in the usual way (cf. Postman et al. 1989; PVJ; PV95), by fitting the cluster density, as a function of $z$ (see the above reference for details). Cluster distances are estimated using the standard relation:

$$R = \frac{c}{H_\circ q_\circ^2 (1+z)} \left[ q_\circ z + (1-q_\circ)(1 - \sqrt{2q_\circ z + 1}) \right] \qquad (15)$$

with $H_\circ = 100\, h$ km sec$^{-1}$ Mpc$^{-1}$ and $q_\circ = \Omega_o/2$. Strictly speaking, eq.(15) holds only for vanishing cosmological constant. Therefore, for a consistent comparison with the simulation models, we should use different $R$–$z$ relations for the Abell/ACO analysis. However, we verified that final results are essentially independent of the choice of the $(\Lambda, \Omega_o)$ parameters used in the simulations. For this reason, in the following we will present results for real data only based on assuming eq.(15) with $q_o = 0.2$.

PVJ and PV95 found that the Abell and ACO cluster number densities, out to their limit of completeness, are $\sim 1.4 \times 10^{-5}\, h^3$ Mpc$^{-3}$ and $\sim 2.1 \times 10^{-5}\, h^3$ Mpc$^{-3}$, corresponding to mean separations $d_{cl} \approx 41\, h^{-1}\, Mpc$ and $d_{cl} \approx 36\, h^{-1}\, Mpc$, respectively. The higher space-density of ACO clusters is partly due to the huge Shapley concentration (Shapley 1930), but a significant part is also due to systematic density differences between the Abell and ACO cluster samples, as a function of $z$, which has been noted in a number of studies (cf. PV91 and references therein) and which could be attributed to the high sensitivity of the IIIa–J emulsion plates. In PV95, this effect was taken into account by normalizing the densities of the two samples using a radial *matching* function, $W(R)$, which is defined as the ratio between the average densities for Abell and ACO clusters at equal volume shells.

In the following, we compare results based on the Abell/ACO sample with those derived from our simulated cluster populations, selected so that $d_{cl} = 40\, h^{-1}\, Mpc$. Variations in $d_{cl}$ of the order of the Abell-ACO difference, does not significantly affect the resulting statistical properties. Finally, following PV95 we restrict our analysis of the real cluster sample within a maximum distance of $R_{max} = 240\, h^{-1}$ Mpc, in order to minimize the uncertainties due to the approximate character of the redshift selection function, $P(z)$, and of the radial *matching* function, $W(R)$, especially at large distances.

## 4 STATISTICS OF THE DISCRETE CLUSTER DISTRIBUTION

Our first statistical test for comparing the real data and the simulations involves the evaluation of the quantity $J_3(R)$, defined by eq. (2). It is straightforward from the definition of this quantity to construct the estimator

$$J_3(R) = \frac{R^3}{3}\left(\frac{N_{nb}}{\overline{N}} - 1\right), \qquad (16)$$

where $N_{nb}$ is the average number of cluster neighbours within a distance $R$ from a cluster, while $\overline{N}$ is the expected number of neighbours for a random cluster distribution (estimated at the positions of the real clusters). Therefore, $J_3(R) \propto R^{3-\gamma}$ as long as $\xi(r) \propto r^{-\gamma}$.

It has been argued (cf. KR94) that the scale at which the power-law shape of $\xi(r)$ breaks and firstly crosses zero, is a potentially powerful test for cosmological models. However, since such a scale corresponds by definition to the weak clustering regime, its detection can be heavily affected by statistical noise; for an explicit demonstration of this, see Paper I. In this respect, the analysis of $J_3$ should have the advantage of being more stable and suffering less from observational biases.

Table 2. $J_3(R)$ values at different scales (in $h^{-1}Mpc$ units) for both simulated and real cluster distributions.

| Model | $J_3(R) \times 10^{-3} (h^{-1}Mpc)^3$ | | |
|---|---|---|---|
| | 28.9 | 43.2 | 64.3 |
| SCDM | $5.00 \pm 0.64$ | $5.55 \pm 1.34$ | $3.86 \pm 2.12$ |
| TCDM | $4.93 \pm 0.74$ | $6.05 \pm 1.60$ | $5.71 \pm 2.74$ |
| LOWH | $5.12 \pm 0.79$ | $6.10 \pm 1.60$ | $5.16 \pm 2.83$ |
| CHDM | $8.39 \pm 0.94$ | $10.19 \pm 1.98$ | $8.77 \pm 3.36$ |
| $\Lambda CDM_1$ | $8.47 \pm 1.06$ | $11.04 \pm 2.23$ | $11.67 \pm 4.33$ |
| $\Lambda CDM_2$ | $18.41 \pm 1.96$ | $23.88 \pm 3.03$ | $26.36 \pm 6.91$ |
| Abell/ACO | $8.10 \pm 1.69$ | $10.28 \pm 2.53$ | $6.96 \pm 5.17$ |

In Figure 7 we plot $J_3(R)$ for the real data (filled circles) and simulations (open circles). We estimated $\overline{N}$ for the real data by averaging over 100 random samples, having the same selection criteria (boundaries, galactic extinction function, redshift selection and systematic Abell/ACO differences) as the real one. Error bars for the simulated samples are $1\sigma$ scatter over the ensemble of 50 realizations. In Table 2 we report values of $J_3$ for data and simulations at three different scales. The quoted uncertainties for real cluster analysis are $1\sigma$ scatter estimated over an ensemble of 100 bootstrap resamplings. Such errors are not plotted in Figure 7. In fact, since we are asking which is the probability that a given model generates a result like that of the observed cluster distribution, its 'success' is just measured by the distance of the real data point from the cosmic r.m.s. error bars. This should be taken into account when judging to which confidence level a model has to be accepted or rejected. In Paper II we verified that intermediate scales of few tens of Mpcs are best suited to constrain DM models, when using the cluster distribution, smaller and larger scales being affected by shot-noise and low signal-to-noise ratio, respectively. Furthermore, richness contamination of cluster correlations should have effects only on rather small scales, below $10\,h^{-1}Mpc$ (see Olivier et al. 1993). For these reasons, in the present analysis we do not consider scales much smaller than $20\,h^{-1}Mpc$ as well as larger than $60\,h^{-1}Mpc$.

For the Abell/ACO sample, $J_3(R)$ increases up to $R \simeq 35\,h^{-1}Mpc$, flattens at a scale corresponding to the break of the power-law shape of $\xi(r)$, and eventually declines at $R \gtrsim 50\,h^{-1}Mpc$, after which $\xi(r)$ becomes negative. By comparing this result with those of the simulations, it turns out that the only models which overcome this test are CHDM and $\Lambda CDM_1$, although both of them seem to produce too strong clustering at smal scale and the second gives a marginal clustering excess at the largest scale. All the other models are ruled out at a $> 2\sigma$ level. The cluster distributions for the SCDM, LOWH and TCDM models are too weakly clustered over the whole scale range. Note that the SCDM has a rather flat $J_3(R)$ profile, according to the expectation that this model has a cluster two-point correlation function which declines rapidly beyond $\sim 20\,h^{-1}Mpc$. Conversely, $\Lambda CDM_2$ generates too much clustering, with $J_3(R)$ increasing up to $R \gtrsim 60\,h^{-1}Mpc$. However, we should keep in mind that the large clustering detected in this model ought to be an overestimate, in the light of the considerations discussed in Section 2.2. Therefore, if we follow the suggestion that cluster correlations are nearly independent of $\sigma_8$ (cf. Croft & Efstathiou 1994; see also Section 2.2), and we take the results of $\Lambda CDM_1$ at $\sigma_8 = 0.8$ to be representative of those at $\sigma_8 = 1.3$, we can conclude that the $\Lambda CDM$ model is viable as far as cluster clustering is concerned (see also Dalton et al. 1994; Gaztañaga et al. 1995). Note that the results for TCDM and LOWH are remarkably similar. This agrees with the expectation that, as far as the shape of the power-spectrum is concerned, a change in the Hubble parameter $h$ is roughly equivalent to a change in the spectral index $n$ according to the relation $\Delta h = -\Delta n$ (cf. Lyth & Liddle 1994).

## 5 STATISTICS OF THE SMOOTHED CLUSTER DISTRIBUTION

In order to facilitate the comparison of our results with those obtained by PV95 from the combined Abell/ACO cluster sample, we followed basically the same procedure as they did, and which we briefly describe below. We obtain a continuous cluster density field by smoothing the cluster distribution on a grid, with grid-cell width of $20\,h^{-1}$ Mpc ($16^3$ grid-points), using a Gaussian kernel:

$$\mathcal{W}(|\mathbf{x}_i - \mathbf{x}_g|) = \left(2\pi R_{sm}^2\right)^{-3/2} \exp\left(-\frac{|\mathbf{x}_i - \mathbf{x}_g|^2}{2 R_{sm}^2}\right). \quad (17)$$

The smoothed cluster density, at the grid-cell positions $\mathbf{x}_g$, is then:

$$\rho(\mathbf{x}_g) = \frac{\sum_i \rho(\mathbf{x}_i)\mathcal{W}(|\mathbf{x}_i - \mathbf{x}_g|)}{\int \mathcal{W}(|\mathbf{x} - \mathbf{x}_g|)d^3x}, \quad (18)$$

where the sum is over the distribution of clusters with positions $\mathbf{x}_i$. In order to study the cluster density field at different smoothing scales, we use three radii for the Gaussian kernel: $R_{sm} = 20, 30$ and $40\,h^{-1}$ Mpc with $|\mathbf{x}_i - \mathbf{x}_g| \le 3R_{sm}$. Therefore the integral in the denominator of eq. (18) has a value smaller than unity ($\simeq 0.97$).

### 5.1 The probability density function

As a first test for the smoothed cluster density field, we work out the probability density function, $f(\varrho)$, which represents a low-order (one-point) statistics. We then compare the pdf of each set of cluster simulations with the observed Abell/ACO pdf, derived by PV95, as well as with the following theoretical models.

(a) The Gaussian distribution given by

$$f(\varrho) = \frac{1}{\sqrt{2\pi\sigma^2}} \exp\left[-\frac{(\varrho-1)^2}{2\sigma^2}\right], \quad (19)$$

where $\sigma$ is the standard deviation of $\varrho$ ($\equiv \rho/\langle\rho\rangle$). If $f(\varrho)$ is a Gaussian then it should be defined in an infinite interval, which implies that $f(\varrho < 0) \ne 0$. Since, however $\varrho \ge 0$ by definition, $f(\varrho)$ is expected to be well approximated by a Gaussian only in the limit $\sigma \to 0$. In this case the skewness, $\gamma$ ($\equiv \langle\delta^3\rangle$), vanishes. Even in the case of an initial Gaussian density field, the gravitational evolution acts in such a way as to increase the variance $\sigma^2$, and thus, due to the constraint $\varrho \geq 0$, $f(\varrho)$ has to become positively skewed. For as long as the variance $\sigma^2 \equiv \langle \delta^2 \rangle$ is small, the deviation of the pdf shape from a Gaussian is well approximated by the Edgeworth expansion (Colombi 1994).

(b) The lognormal distribution given by

$$f(\varrho) = \frac{1}{\sqrt{2\pi\sigma_L^2}} \exp\left[-\frac{(\ln\varrho - \mu_L)^2}{2\sigma_L^2}\right] \frac{1}{\varrho}, \qquad (20)$$

where $\varrho$ is obtained through an exponential transformation of a Gaussian random variable $\chi$ as $\varrho = \exp(\chi)$. In eq. (20), $\mu_L$ and $\sigma_L$ are the mean and standard deviation of $\ln\varrho$ respectively. It has been argued that this distribution describes the distribution of density perturbations resulting from Gaussian initial conditions in the weakly non–linear regime Coles & Jones 1991). Bernardeau & Kofman (1994) have shown that the lognormal distribution is **not** really a natural consequence of mildly non–linear gravitational evolution, but a very convenient fit only in some portion of the $(\sigma, n)$-plane (i.e. $\sigma \ll 1$ and spectral index $n \approx -1$). It has nevertheless been found to give an extremely good fit to the CDM density and the IRAS galaxy pdf in the weakly–linear regime (Kofman et al. 1994), as well as to the observed Abell/ACO cluster distribution (PV95).

(c) The pdf resulting from the application of the ZA to Gaussian initial fluctuations (Kofman et al. 1994):

$$\begin{aligned}
f(\varrho) &= \frac{9 \times 5^{3/2}}{4\pi N_s \varrho^3 \sigma^4} \\
&\times \int_{3\varrho^{-1/3}}^{\infty} ds\, e^{-(s-3)^2/2\sigma^2} \left(1 + e^{-6s/\sigma^2}\right) \\
&\times \left(e^{-\beta_1^2/2\sigma^2} + e^{-\beta_2^2/2\sigma^2} - e^{-\beta_3^2/2\sigma^2}\right); \\
\beta_n(s) &= \sqrt{5}\, s\, \{1/2\, + \cos[2/3\, (n-1)\pi \\
&\quad + 1/3\, \arccos\left(54/\varrho s^3 - 1\right)]\},
\end{aligned} \qquad (21)$$

where $\sigma$ is the rms amplitude of density fluctuations and $N_s$ is the average stream number per Eulerian point.

Note that discreteness effects could be important since they affect the shape of the pdf and the estimation of its moments; especially at small $R_{sm}$, when the number of clusters in the Gaussian sphere is small, and/or when the smoothing fails to create a continuous density field due to discreteness (in our case this is apparent in the $R_{sm} = 20\ h^{-1}$ Mpc case for $\varrho \leq 0.8$). In the case of a Poisson sampling of an underlying continuous density field, the shot–noise contributions to the moments can be easily estimated and corrected for (cf. Peebles 1980). The cluster distribution cannot, however, be meaningfully regarded as a Poisson sampling of the underlying (galaxy) distribution, since clusters are expected to form only at high density peaks. Consequently the Poissonian shot–noise correction could not give a reasonable description of discreteness effects (Coles & Frenk 1991; Borgani et al. 1994). Moreover, Gaztañaga & Yokoyama (1993) have shown that the smoothing process itself considerably suppresses these shot–noise effects. For these reasons PV95 did not use any shot–noise corrections. To make a consistent comparison of our models with the data, we also did not include such corrections in our analysis. Since all the model cluster distributions have the same mean number density and we treat them similarly, the possible effects of shot–noise are accounted for in the same way in both the data and the simulations: we are therefore comparing like with like.

In Table 3 we present the results of the comparison between the simulation pdf, the PV95 Abell/ACO cluster pdf, the lognormal distributions and a Gaussian distribution, using a $\chi^2$–test defined as:

$$\chi^2 = \sum_i^{bins} \left(\frac{f_i^{sim}(\varrho) - f_i^{theor}(\varrho)}{\epsilon_i}\right)^2, \qquad (22)$$

where the weights $\epsilon_i^2$ correspond to cosmic variance. Note that we derived the simulation pdf in redshift space so that a consistent comparison with the PV95 results can be made; the comparison with theoretical models is done in real space.

In Figure 8 we present the simulation cluster pdfs for the CHDM and SCDM models together with the PV95 Abell/ACO pdf at $R_{sm} = 20$ and $40\, h^{-1} Mpc$, and in Figure 9 we make the comparison between these two cosmological models and the theoretical distributions. The error bars represent the scatter around the ensemble mean values (cosmic variance). As in Table 3, comparisons with theoretical models and real data are made in real space and in redshift space, respectively. There is an excellent agreement between the CHDM and the Abell/ACO cluster pdfs while in the SCDM case there is a clear discrepancy at small ($\leq 0.5$) and large $\varrho$'s.

It is apparent that:

(i) The Gaussian distribution of does not provide a good fit at any $R_{sm} \leq 30\ h^{-1}$ Mpc and for any model (cf. column 4 of Table 3). For $R_{sm} = 40\ h^{-1}$ Mpc the Gaussian fit is acceptable only for the SCDM, LOWH and $\Lambda$CDM$_2$ models.

(ii) The Zel'dovich pdf model of eq.(21) is inconsistent with the simulation results, even though the underlying dynamics governing the cluster distribution are described by the ZA. We find that this distribution is ruled out at a confidence level larger than 99.99%, for $R_{sm} \leq 30\ h^{-1}$ Mpc for all the simulation models. This is the reason why we did not show results for this model in Table 3. The SCDM, TCDM, CHDM and $\Lambda$CDM$_1$ models are only consistent at a $\sim 20\% - 25\%$ level for $R_{sm} = 40\ h^{-1}$ Mpc. One may argue that, since the ZA pdf is designed to describe the DM clustering, its failure for the cluster distribution is nothing but the consequence of not accounting for the mass within clusters in the analysis of their distribution. To check this, we repeated the analysis by weighting simulated clusters according to their mass and found no appreciable differences in the pdf shapes.

(iii) All simulation pdfs are well approximated by lognormal distribution of eq.(20), irrespective of their different power–spectra, especially when the larger smoothing radii are considered (cf. column 3 of Table 3). Note that at the $R_{sm} = 20\ h^{-1}$ Mpc case the comparison is done for $\varrho > 0.8$ because, at lower values of $\varrho$, discreteness effects introduce significant noise. Therefore, in contrast to the case of the matter distribution, the lognormal fit to the cluster pdf is more likely to be connected with the high–peak biasing description of cluster formation than being due to non–linear gravitational effects, which dominate much smaller scales (Bernardeau & Kofman 1994).

**Table 3.** $\chi^2$ probabilities that the indicated simulation model pdf could have been drawn from a parent distribution given by the lognormal, Gaussian or the real cluster (PV95) pdf. For the $R_{sm} = 20\,h^{-1}$ Mpc case the comparison is performed for $\varrho > 0.8$ (see text). Also reported in Column 6 are the values of the reduced skewness $S_3$ for simulations and for real data.

| Model | $R_{sm}$ | $P_{\chi^2}^{LN}$ | $P_{\chi^2}^{G}$ | $P_{\chi^2}^{data}$ | $S_3$ |
|---|---|---|---|---|---|
| SCDM | 20 | 0.02 | 0.00 | 0.01 | $1.87 \pm 0.29$ |
|  | 30 | 0.83 | 0.00 | 0.75 | $1.83 \pm 0.70$ |
|  | 40 | 0.99 | 0.92 | 0.97 | $1.60 \pm 1.25$ |
| TCDM | 20 | 0.15 | 0.00 | 0.00 | $1.96 \pm 0.32$ |
|  | 30 | 0.99 | 0.00 | 0.99 | $2.02 \pm 0.80$ |
|  | 40 | 0.99 | 0.06 | 0.99 | $1.96 \pm 1.44$ |
| LOWH | 20 | 0.02 | 0.00 | 0.00 | $1.87 \pm 0.28$ |
|  | 30 | 0.89 | 0.00 | 0.97 | $1.87 \pm 0.69$ |
|  | 40 | 0.99 | 0.49 | 0.97 | $1.75 \pm 1.18$ |
| CHDM | 20 | 0.10 | 0.00 | 0.95 | $1.93 \pm 0.30$ |
|  | 30 | 0.99 | 0.00 | 0.98 | $1.96 \pm 0.65$ |
|  | 40 | 0.99 | 0.00 | 0.63 | $1.81 \pm 1.08$ |
| $\Lambda$CDM$_1$ | 20 | 0.39 | 0.00 | 0.98 | $1.91 \pm 0.26$ |
|  | 30 | 0.33 | 0.00 | 0.70 | $1.96 \pm 0.53$ |
|  | 40 | 0.02 | 0.72 | 0.22 | $1.94 \pm 0.89$ |
| $\Lambda$CDM$_2$ | 20 | 0.07 | 0.00 | 0.00 | $1.94 \pm 0.23$ |
|  | 30 | 0.83 | 0.00 | 0.00 | $2.00 \pm 0.49$ |
|  | 40 | 0.99 | 0.00 | 0.00 | $1.95 \pm 0.80$ |
| Abell/ACO | 20 | 0.00 | 0.00 | – | $1.81 \pm 0.23$ |
|  | 30 | 0.06 | 0.00 | – | $1.78 \pm 1.30$ |
|  | 40 | 0.99 | 0.27 | – | $1.76 \pm 1.85$ |

(iv) The scale which best discriminates between different models and the Abell/ACO data is clearly $R_{sm} = 20\,h^{-1}\,Mpc$ (cf. column 5 of Table 3). At larger scales, all the models, except $\Lambda$CDM$_2$, produce acceptable fits. The only models that produce a pdf consistent with the PV95 results, at all 3 smoothing radii, are CHDM and $\Lambda$CDM$_1$, with the former performing however systematically better.

### 5.2  Moments of the pdf

According to eq. (3), the moments of the pdf give a large weight to the high density tail ($\delta > 1$) of the pdf. They are therefore expected to suffer less from shot–noise effects, which are smaller in the overdense parts of the distribution. In Paper II we presented results about the variance, $\sigma^2 = \langle \delta^2 \rangle$, and the skewness, $\gamma = \langle \delta^3 \rangle$.

It has been argued on several grounds (e.g. Coles & Frenk 1991) that the relation

$$\gamma \approx S_3 \left( \sigma^2 \right)^2, \qquad (23)$$

with $S_3$ nearly independent of scale, should describe the clustering of cosmic structures. Although at small scales, below a few Mpc, eq. (23) is predicted by models of non–linear gravitational clustering (e.g. Borgani 1995 and references therein), at the larger scales, sampled by galaxy clusters, it is expected to hold due to mildly non–linear evolution as well as by the bias relating the cluster and DM distributions. The resulting $S_3$ values at different $R_{sm}$ are reported in column 6 of Table 3 for both the simulations and the Abell/ACO sample. We note that (a) the reduced skewness $S_3$ is always independent of the scale with a good accuracy, and (b) it takes the same value $S_3 \simeq 1.9$ for all the models, within statistical fluctuations, and consistent with the observational data (PV95).

Accordingly, we conclude that, for the cluster distribution, only the amplitude of clustering, and not its *nature*, depends on the initial power–spectrum. This suggests that both the lognormal pdf shape and the $S_3$ value observed for real data are natural consequences of high–peak biasing and possibly of the random–phase assumption of the primordial density field.

### 5.3  The $\beta$–parameter of clusters

An interesting quantity, which relates the large–scale clustering to the linear peculiar velocity field of clusters is the $\beta$–parameter, defined as

$$\beta_{cl} = \frac{f(\Omega_\circ)}{b_{cl}}, \qquad (24)$$

where $f(\Omega_\circ) \simeq \Omega_\circ^{0.6}$ to a good accuracy (e.g. Peebles 1993) is the linear velocity factor and

$$b_{cl} = \frac{\sigma_{cl}}{\sigma_{DM}} \qquad (25)$$

is the biasing parameter, here defined as the ratio between the r.m.s. fluctuations for the cluster and the DM density fields.

For the simulations we directly estimate $\beta_{cl}$, since we know *a priori* $\Omega_\circ$ and we can compute $\sigma_{DM}$ for each model according to

$$\sigma_{DM}(R_{sm}) = \left[ \frac{1}{2\pi^2} \int dk\, k^2\, P(k)\, \mathcal{W}_{R_{sm}}^2(k) \right]^{1/2}, \qquad (26)$$

where $\mathcal{W}_{R_{sm}}(k) = \exp(-k^2 R_{sm}^2/2)$ is the Fourier transform of the window function of eq. (17).

¿From the observational size, $\beta_{cl}$ can be estimated by comparing the linear velocity induced by the cluster distribution on the Local Group (LG) with the LG velocity as measured from the CMB temperature dipole. Results of these analyses (PV91; Scaramella, Vettolani & Zamorani 1991; Scaramella 1995; Branchini & Plionis 1995; Tini Brunozzi et al. 1995; Plionis 1995) consistently indicate that $\beta_{cl} = 0.20 \pm 0.05$. From their power–spectrum analysis, Jing & Valdarnini (1993) and Peacock & Dodds (1994) verified that at least the relative biasing between clusters and optically as well as infrared selected galaxies is independent of the scale to a quite good accuracy. In particular, Peacock & Dodds (1994) found that $b_{cl}/b_{IRAS} \simeq 4.5$, for the relative biasing between clusters and IRAS galaxies. Therefore, if $\beta_{IRAS} = 1.0 \pm 0.2$ (Peacock & Dodds 1994), it turns out that $\beta_{cl} = 0.22 \pm 0.05$, in agreement with the other value.

In Figure 10 we plot the $\beta_{cl}$ parameters for the cluster simulations at different $R_{sm}$. For sake of clarity, we plot $1\sigma$ error bars, due to cosmic variance, only for LOWH. The heavy–dashed horizontal lines delineates the $1\sigma$ band for the observational result, $\beta_{cl} = 020 \pm 0.05$. Two main conclusions can be drawn from this figure.

(i) Independent of the model considered, $\beta_{cl}$ is fairly constant over the whole scale range, thus implying a nearly scale–independence also for $b_{cl}$. This result is quite different from the analogous one presented in Paper I, which showed a decreasing trend for $b_{cl}$ at small scales. This confirms how important is the increased resolution and the optimization of the ZA for the reliability of our simulations. ¿From one hand, such a linearity of the biasing is a rather remarkable result, since both the evolution of the density field and the selection of clusters as high-density peaks represent definitely non-linear transformations of the initial fluctuations. From the other hand, this result supports the usual assumption of $b_{cl} = const$, used to infer the shape of the primordial power–spectrum from that of clusters.

(ii) Within the models with $\Omega_o = 1$, the TCDM is the only one that is only marginally consistent with the observational $\beta_{cl}$ value, while all the other three models are perfectly consistent. On the other hand, both the $\Lambda$CDM models give too small values of $\beta_{cl}$. For $\Lambda$CDM$_1$, the small $\beta_{cl}$ is due to the low spectrum normalization, which corresponds to a large $b_{cl}$. Note, however, that for $\Lambda$CDM$_2$ the $\beta_{cl}$ parameter is underestimated, due to the overestimate of the cluster correlations. In fact, if we would allow for $\Lambda$CDM$_2$ clusters to have the same clustering as $\Lambda$CDM$_1$ clusters, then the $\beta_{cl}$ parameter would increase from $\beta_{cl} \simeq 0.11$ to $\beta_{cl} \simeq 0.18$, which is within the observational uncertainties. This indicates that, for a $\Lambda$CDM model, a normalization as high as $\sigma_8 = 1.3$ is required not only to match COBE data and cluster abundances, but also to generate a high enough $\beta_{cl}$ or, equivalently, to generate large enough peculiar velocities.

## 6 CONCLUSIONS

In this paper we have compared results of a statistical analyses of our simulated cluster distributions and of a combined sample of Abell/ACO clusters. Our cluster simulations, which are based on the truncated Zel'dovich approximation (TZA), are extremely cheap computationally; each realization takes about 6 minutes of CPU on a HP735/125 workstation. This has allowed us to run a large number of realizations (50) for each model, so as to properly estimate the cosmic variance.

In order to check the reliability of the TZA for cluster simulations, we compared them with analogous results based on PM N-body simulations, having the same initial conditions, mass resolution and cluster identification criteria as the TZA simulations. We find that, as long as clustering is only mildly non-linear on the cluster mass scale (i.e. $\sigma_8 \lesssim 1$), the TZA reproduces N-body results with remarkable accuracy. The agreement between the corresponding cluster distributions occurs not only in a statistical sense;

**Table 4.** Summarizing scheme of the tests applied to the considered DM models. The null score means that the test rejects the model at least at the $2\sigma$ level.

| Model | $N(>M)$ | $J_3$ | $pdf$ | $S_3$ | $\beta_{cl}$ |
|---|---|---|---|---|---|
| SCDM | 0 | 0 | 0 | 1 | 1 |
| TCDM | 0 | 0 | 0 | 1 | 1 |
| LOWH | 1 | 0 | 0 | 1 | 1 |
| CHDM | 1 | 1 | 1 | 1 | 1 |
| $\Lambda$CDM | 1 | 1 | 1 | 1 | 1 |

cluster positions for TZA and PM simulations are shown to be extremely close *point–by–point*. Furthermore, comparing the resulting two–point correlation function for our CHDM clusters with that obtained by Klypin & Rhee (1994) from their higher resolution PM simulations of the same model, we find a extremely good agreement, even down to quite small scales ($\simeq 7 h^{-1} Mpc$), at which one can doubt the validity of the TZA. The reason for this remarkable success of the TZA in reproducing N-body results lies in its ability to account for non–local (i.e. long–wavelength) effects when moving particles from their initial (linear) positions to their correct evolved ones (cf. Pauls & Melott 1994; Sathyaprakash et al. 1994).

However, if large spectrum normalizations are considered ($\sigma_8 \gtrsim 1$), non–linear gravitational effects, like infall and merging of structures starts playing a role in N-body simulations, while they are not accounted for in the TZA. As a result, the cluster clustering is stable for N-body simulations, while it increases with $\sigma_8$ in TZA simulations. We also verified that no substantial changes are found by going to the second–order in Lagrangian perturbative theory. This suggests the non-perturbative nature of effects like merging or infall.

We applied applied five different tests to the considered DM models: (a) cluster abundances; (b) cluster correlations ($J_3(R)$); (c) shape of the cluster probability density function; (d) reduced skewness $S_3$; (e) determination of the $\beta_{cl}$ parameter. In Table 6 we summarize the results. Failures, indicated with a '0', represent results incompatible with observations at the $2\sigma$ level. For the $\Lambda$CDM model we report the results of the COBE–normalized version ($\Lambda$CDM$_2$; $\sigma_8 = 1.3$) only for cluster abundances. Results concerning the clustering analyses are for the low–normalization version ($\Lambda$CDM$_1$; $\sigma_8 = 0.8$), which are expected to be similar to those at $\sigma_8 = 1.3$ on the basis of stable clustering arguments, suggested by N-body simulations.

As for the clustering analysis, we have analyzed both, the discrete cluster distribution using the $J_3$ statistic, and the smoothed one using the *pdf* statistic. These analyses provide stringent constraints on the DM models we have considered and the only models that pass all the tests are the Cold + Hot DM scenario and the low–density flat CDM model with the lower normalization ($\sigma_8 = 0.8$). Standard, Tilted and Low–H$_o$ CDM versions do not account for the large-scale clustering of the real cluster sample. The low–density model with the larger normalization ($\sigma_8 = 1.3$) produces much stronger clustering than real clusters on all scales. However, in this case the TZA is expected to overestimate the cluster correlations. If, based on the clustering stability

12  S. Borgani, M. Plionis, P. Coles and L. Moscardinisuggested by N-body simulations, we accept that results at $\sigma_8 = 0.8$ are representative of those at $\sigma_8 = 1.3$, we could conclude that $\Lambda$CDM agrees quite well with the data for the clustering tests we have used.

We find that the shape of the probability density function (pdf) is not well approximated by a Gaussian distribution, even at the largest smoothing scale considered. The pdf for both simulations and Abell/ACO clusters is always much better reproduced by the lognormal model than by the Zel'dovich prediction, despite the fact that the ZA governs the underlying dynamics. This shows that, at least for clusters, the lognormal shape of the pdf does not occur by chance, for a limited set of initial conditions, as argued to happen for the galaxy distribution (Bernardeau & Kofman 1994). Instead it is much more likely to be related to the fact that clusters trace the high density peaks of the underlying matter field and, perhaps, to the initial random–phase assumption. Using a $\chi^2$–test to compare the shape of the pdf for both data and simulations, we find the best models to be CHDM and $\Lambda$CDM, consistent with the analysis of $J_3$.

Although the variance and skewness of the smoothed cluster pdf are powerful discriminators of different models (see Paper II), the reduced skewness, $S_3 = \gamma/\sigma^4$, turns out to be independent of the initial spectrum. We always find $S_3 \simeq 1.9$, almost independent of the scale, and consistent with the observational results (Plionis & Valdarnini 1995). One could be tempted to conclude that such a value of $S_3$ is naturally produced by the high-peak selection of clusters, probably combined with the Gaussian nature of the initial fluctuations. Whether the analysis of the reduced skewness represents a test of Gaussian vs. non-Gaussian initial conditions remains to be seen (see also Coles et al. 1993).

By computing the scale–dependence of the parameter $\beta_{cl} = \Omega_o^{0.6}/b_{cl}$, we have verified that the linear biasing prescription used to relate cluster and DM distributions is always satisfied to a good precision. The resulting value of $\beta_{cl}$ depends on the details of the model. Between the $\Omega_o = 1$ models, the only one which can be marginally excluded is the TCDM, whose low normalization ($\sigma_8 = 0.5$) gives rise to a rather low $\beta_{cl}$. Both the two low–density models have a quite low $\beta$–parameter, $\beta_{cl} \simeq 0.1$. However, for $\Lambda$CDM$_2$ this ought to be an underestimate, due to the overestimate of the clustering. On the other hand, taking the cluster correlations measured for $\Lambda$CDM$_1$ and rescaling to the normalization of $\Lambda$CDM$_2$, would give a result perfectly consistent with the observational constraint $\beta_{cl} = 0.20 \pm 0.05$.

As a complementary indication of the success or not of the DM models considered, we compared the predicted cluster abundances with available observational results using the Press & Schechter (1974) approach. The only models that produce adequate abundances are CHDM, the low–density flat CDM model with $\sigma_8 = 1.3$ ($\Lambda$CDM$_2$) and low–$H_o$ CDM. Given the uncertainties in both theory and observations, however, we take these results to be indicative rather than definitive.

The overall picture emerging from these studies is that the large–scale cluster distribution places stringent constraints on models of structure formation. Combining all the above results on clustering and abundances, it emerges that the only models to survive all the constraints we have considered here are CHDM and $\Lambda$CDM (with a marginal preference for the former).

One can also ask whether reasonable modifications of the parameters in the DM models considered above (i.e. the Cold + Hot DM mixture, the values of $\Omega_o$ and $\Omega_\Lambda$ values, the primordial spectral index $n$, the Hubble constant, etc.) could lead to significant changes in the resulting cluster distribution. As an example, it has been shown that slightly decreasing the relative fraction of the hot component in the CHDM model bring it in a much better agreement with the high–redshift detection of collapsed structures (Klypin, Nolthenius & Primack 1995), while whether $\Omega_{hot}$ is shared between one or more neutrino species could also play a role. Verifying which of the CHDM variants produces an acceptable cluster distribution is not a difficult task, thanks to the low computational cost of our simulations.

We believe that, although keeping in mind the limitations of the TZA, in future investigations of specific DM models, the optimal strategy would be, first of all, to run optimal TZA simulations in order to assess the model on large scales. Only after that one should decide whether a model is worth exploring at smaller scales by means of high–resolution, computationally expensive N–body simulations.

### Acknowledgments.

We are grateful to the referee, Joel Primack, for useful suggestions in clarifying the presentation of the results. MP acknowledges the receipt of an EC *Human Capital and Mobility* Fellowship. PC acknowledges the receipt of a PPARC Advanced Research Fellowship. SB and LM have been partially supported by Italian MURST. We are also grateful to PPARC for support under the QMW Visitors Programme in Astronomy GR/J 88357.

**Figure Captions**

**Figure 1.** The average number of streams per Eulerian point, $N_s$, as a function of the Gaussian filtering scale $R_f$ for the six different models. The dotted horizontal line, $N_s = 1.1$, delineates the quasi single-stream regime.

**Figure 2.** Comparison between TZA and PM cluster simulations, at $\sigma_8 = 0.67$ (left panels) and $\sigma_8 = 1$ (right panels). In the upper panels the distribution of clusters obtained using the two simulation techniques are compared. Open circles are the TZA cluster positions (for $R_f = 3 \, h^{-1} \, Mpc$) and filled dots give the PM cluster positions. The lower panels compare the cluster count–in–cell variance for PM simulations and for TZA simulations, for three different values of the filtering radius $R_f$.

**Figure 3.** Comparison between the two–point correlation function for our CHDM cluster simulations (open dots) and for the PM N–body results from Klypin & Rhee (1994) based on the same initial spectrum (filled dots). Their results are obtained with PM simulations having $256^3$ grid points on a box of side $200\,h^{-1}\,Mpc$. Our results are the average over 50 realizations, and errors are $1\sigma$ scatter over this ensemble. The KR94 results are an average over 2 realizations and the error bars are quasi–Poissonian estimates.

**Figure 4.** The two–point cluster correlation functions for TZA simulations based on the first–order (filled circles) and second–order (open circles) Lagrangian theory, at $\sigma_8 = 0.67$ (left panel) and $\sigma_8 = 1$ (right panel). The plotted $\xi(r)$ values are the average over 10 realizations and the corresponding error bars are the $1\sigma$ scatter over this ensemble.

**Figure 5.** The cluster distribution (heavy dots) for the CHDM model superimposed on the DM particle distribution in a slice $10\,h^{-1}\,Mpc$ thick for a box of side $640\,h^{-1}\,Mpc$. It is interesting to note that clusters are strongly correlated with the intersection of filaments of the DM distribution.

**Figure 6.** The abundances of clusters with mass $M > 4.2 \times 10^{14} h^{-1} M_\odot$ for the different DM models, as predicted by the Press & Schechter (1974) formalism, as a function of the critical density contrast $\delta_c$. The horizontal lines are the observational results by White et al. (1993; dotted line) and by Biviano et al. (1993; dashed line) for clusters with mass larger than the above value.

**Figure 7.** The $J_3(R)$ integral as a function of the scale $R$ for the simulated (open circles) and real Abell/ACO (filled circles) cluster distributions. Error bars are plotted only for the simulations and correspond to $1\sigma$ scatter over the ensemble of 50 realizations.

**Figure 8.** Comparison between the pdfs for real (filled circles) and simulated (open circles) cluster distributions. Results only for the SCDM and CHDM are shown at $R_{sm} = 20$ and $40\,h^{-1}\,Mpc$. For the simulations, the analysis is realized in redshift space and the error bars correspond to cosmic r.m.s. scatter.

**Figure 9.** Comparison between the pdfs of simulated cluster distributions and the theoretical models. We plot only results for SCDM and CHDM models at $R_{sm} = 20$ and $40\,h^{-1}\,Mpc$. Solid, long–dashed and short–dashed curves correspond to the lognormal, Zel'dovich and Gaussian model, respectively. Error bars are cosmic r.m.s. scatter.

**Figure 10.** Scale dependence of the cluster $\beta$–parameter for the six different models. For reasons of clarity, we plot the corresponding cosmic r.m.s. scatter only for the LOWH model. Similar uncertainties hold also for the other models. The heavy dashed horizontal lines show the range indicated by observational results (see text), $\beta_{cl} = 0.20 \pm 0.05$.

Figure 3

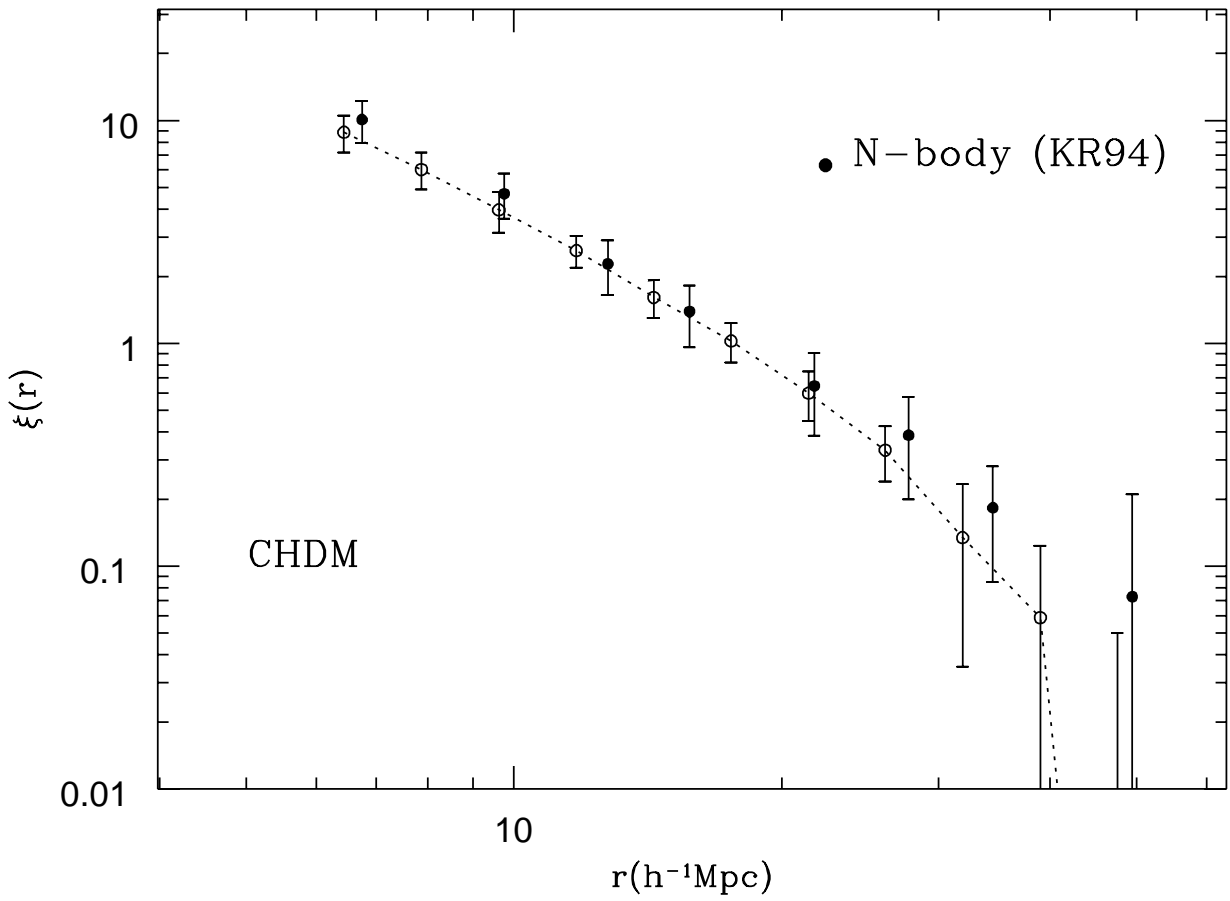

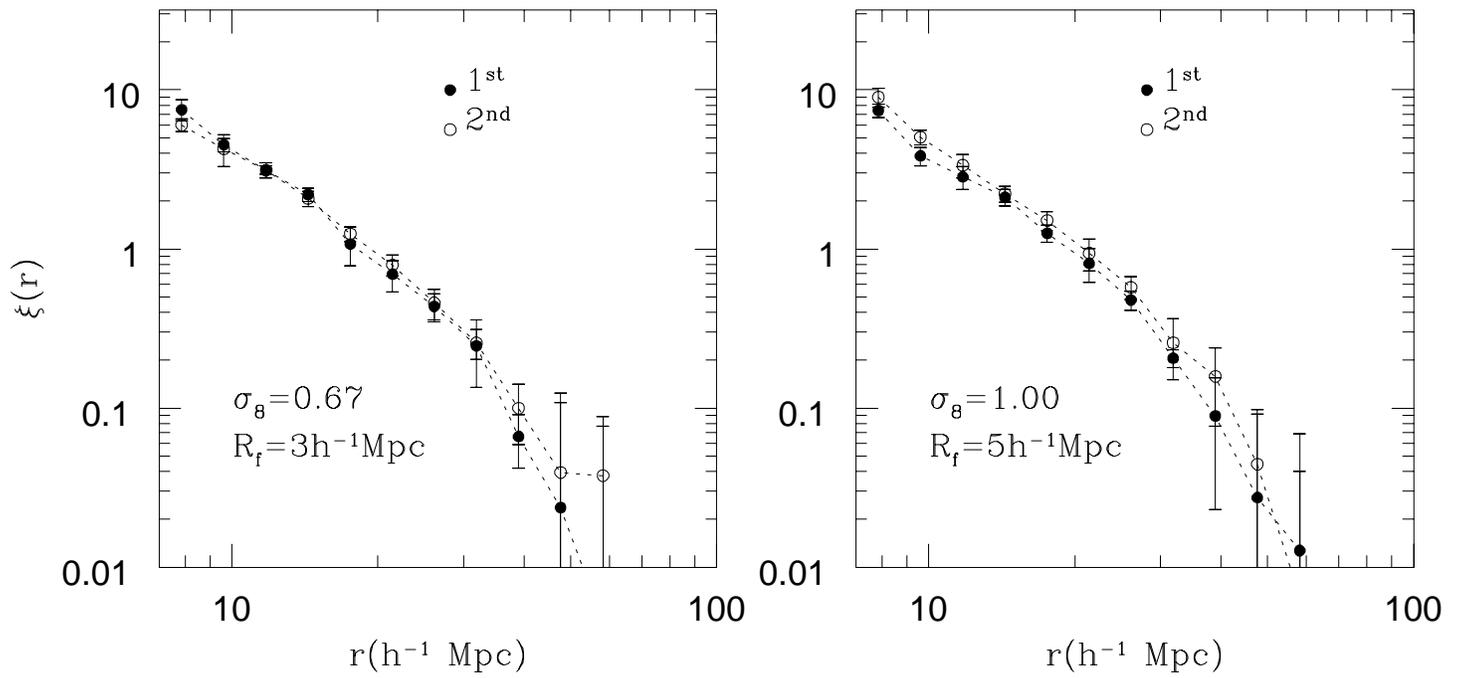

Figure 4

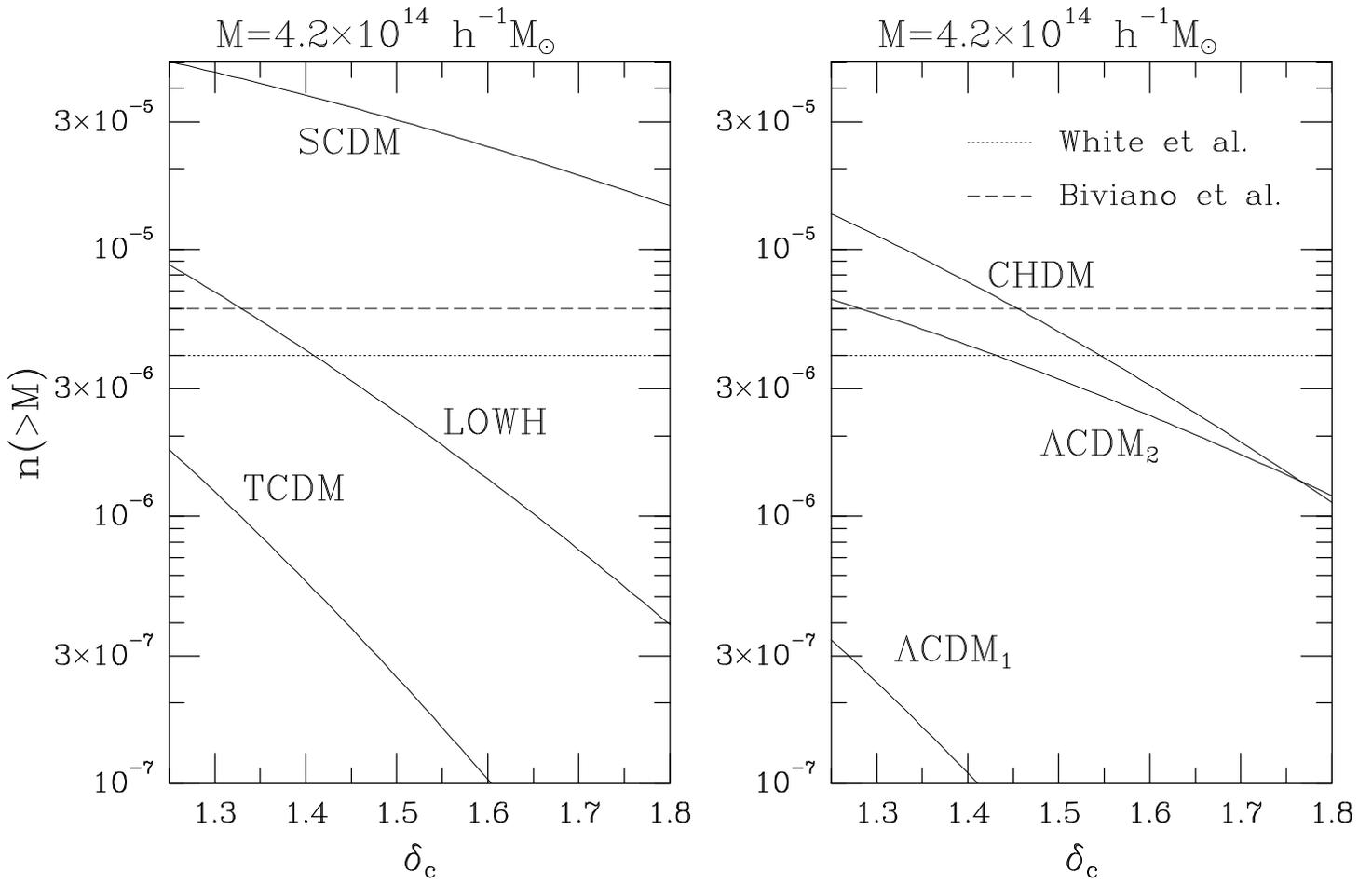

Figure 6

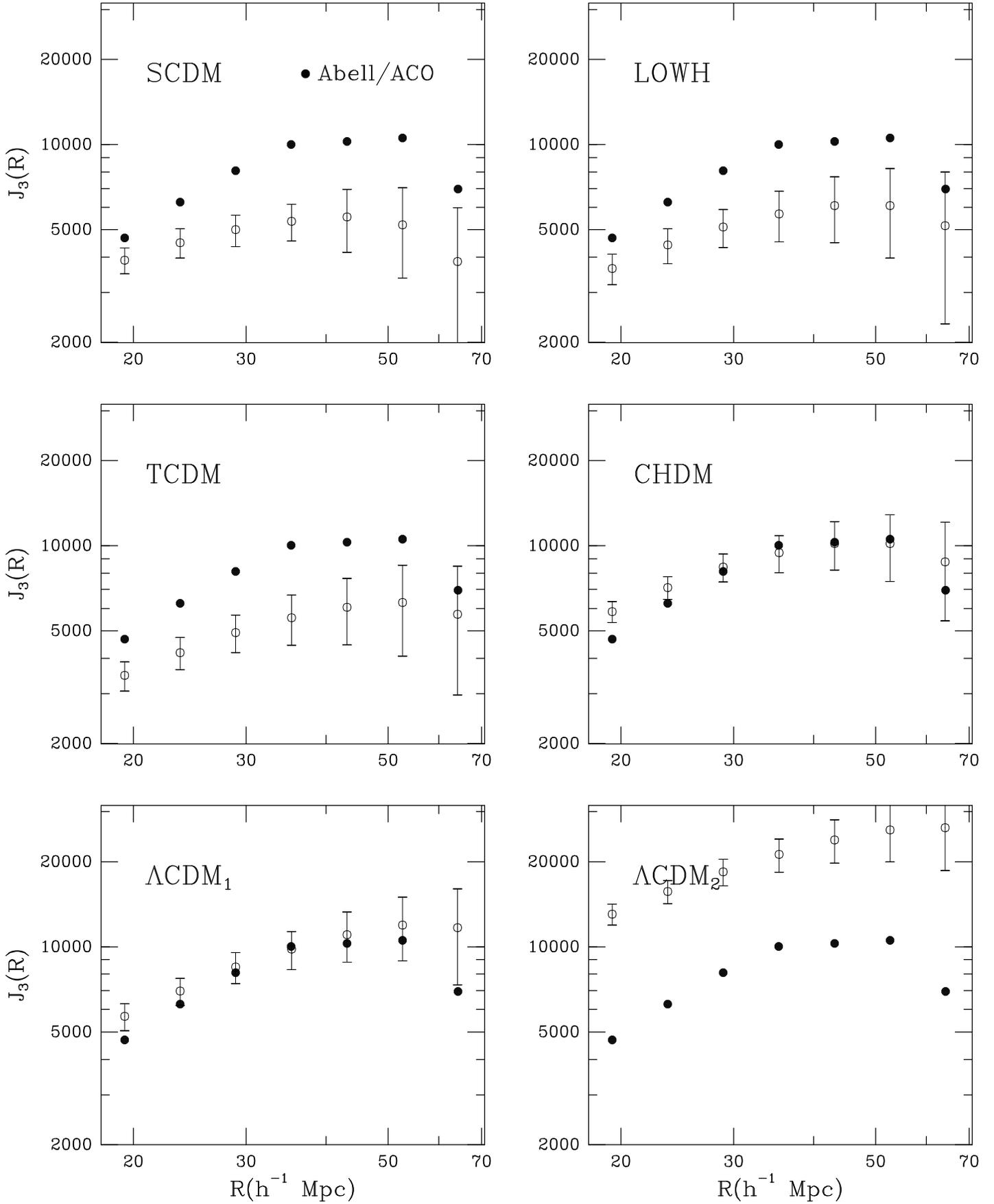

Figure 7

Figure 8

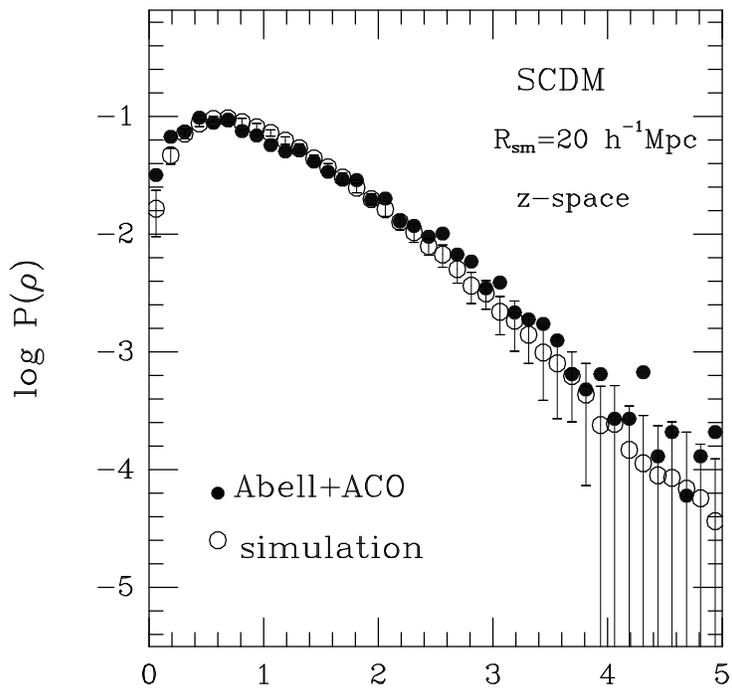
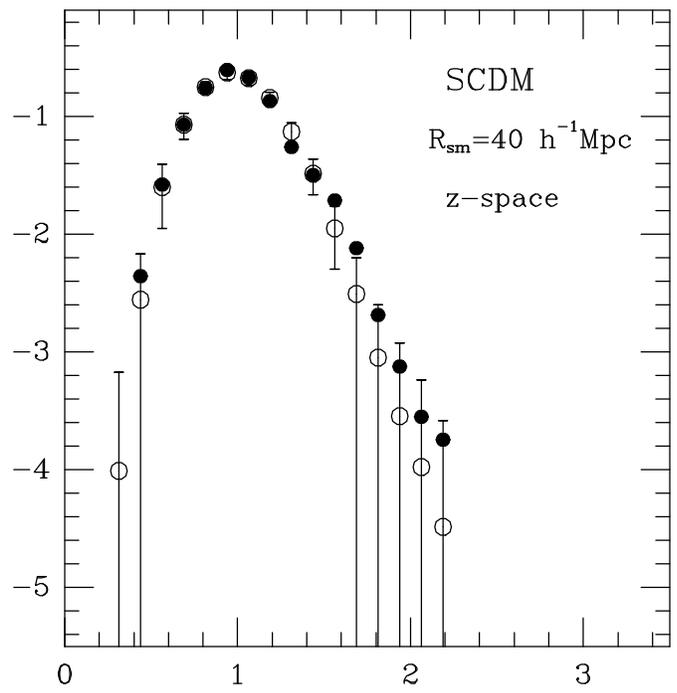
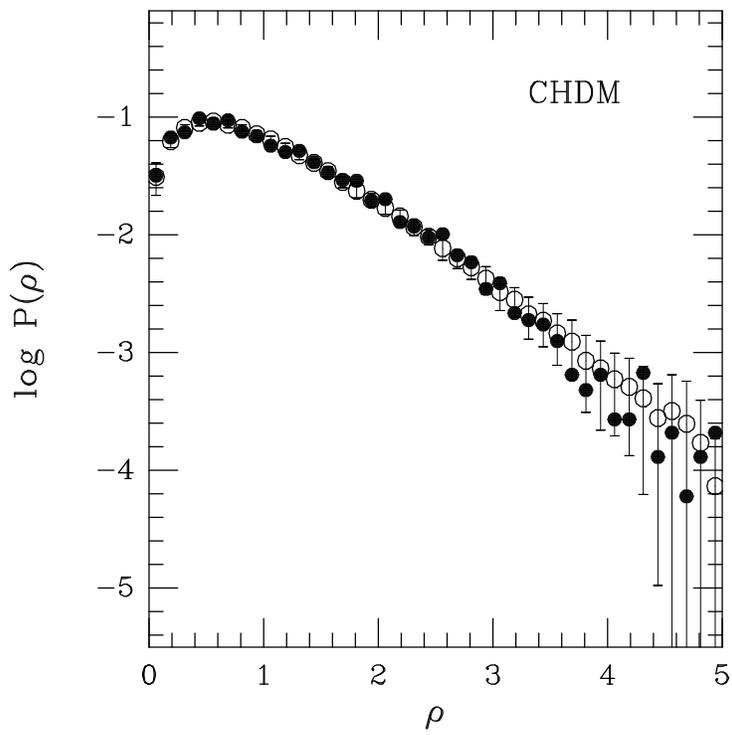
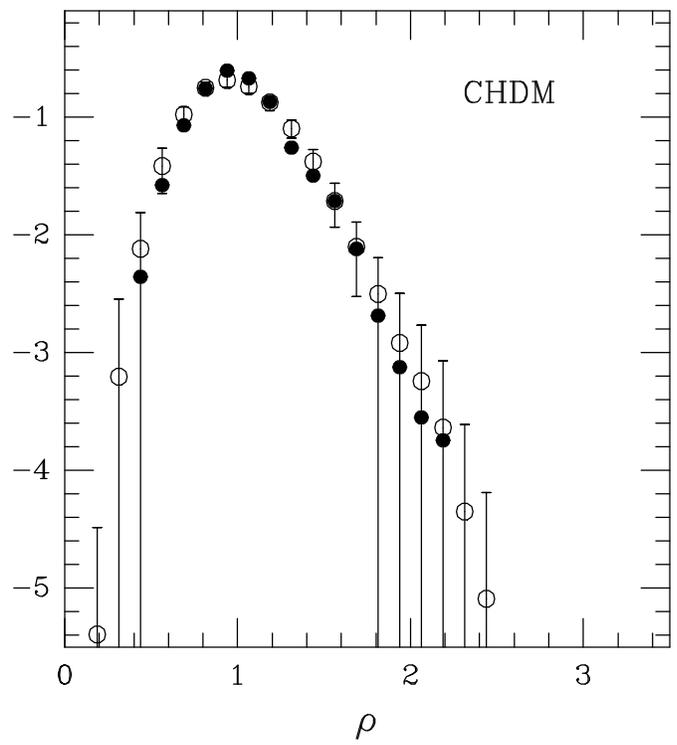

Figure 9

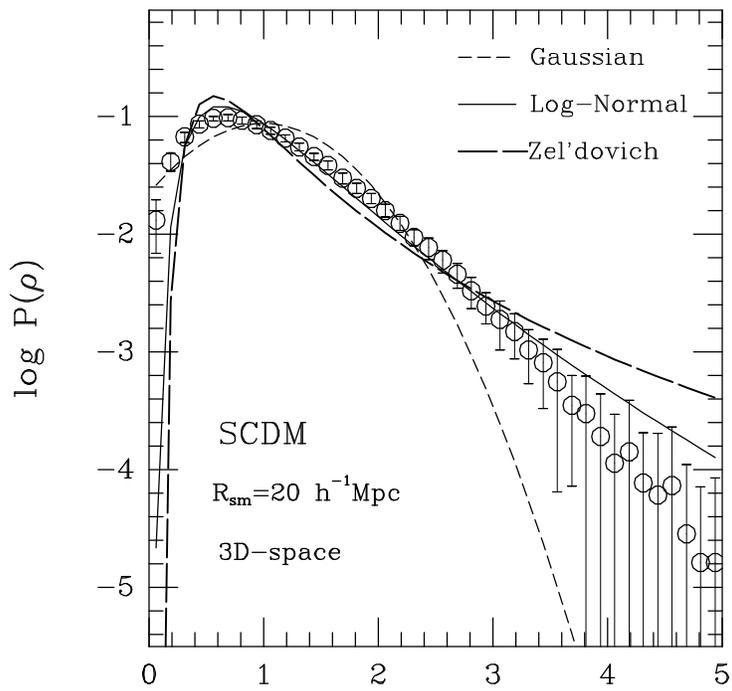
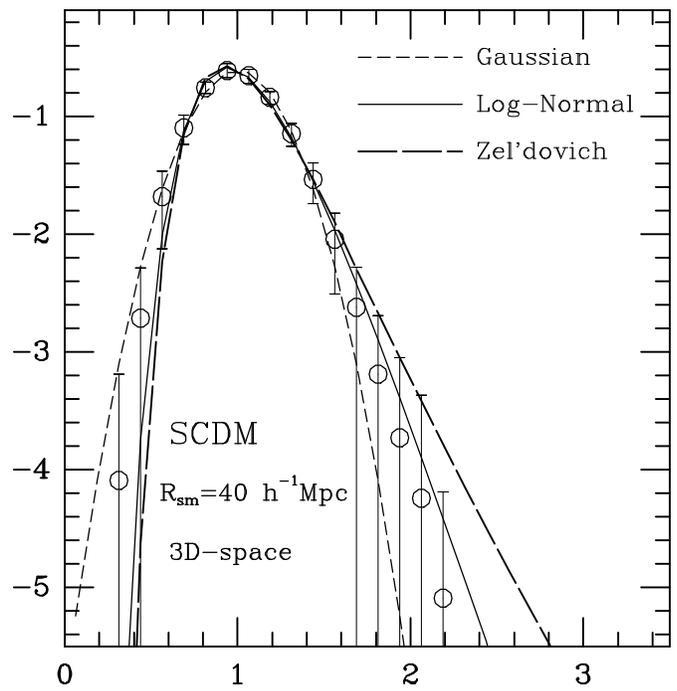
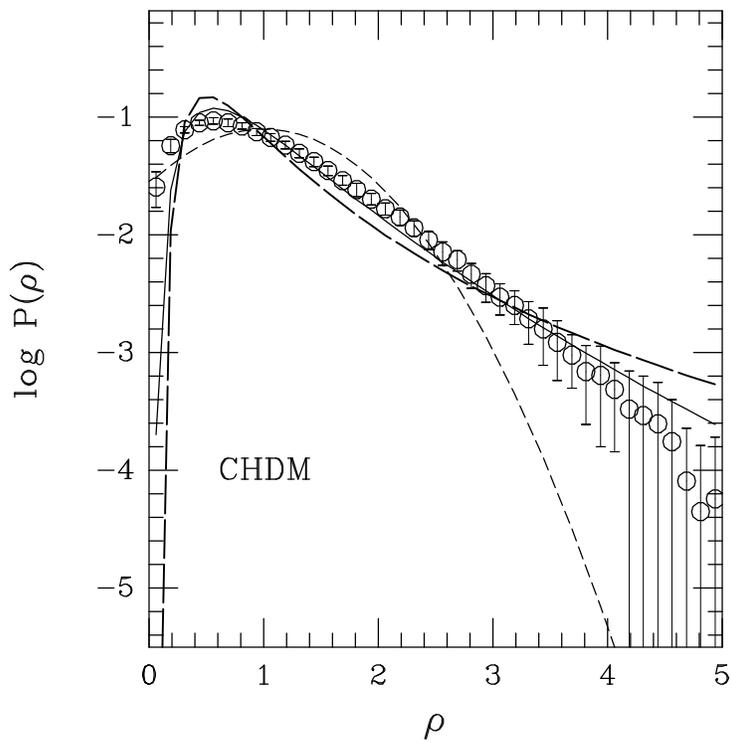
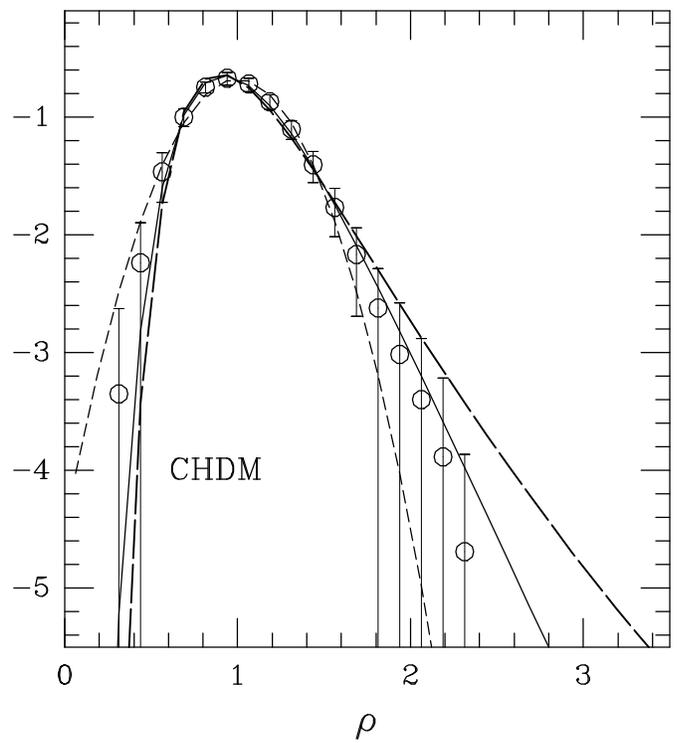

Figure 10

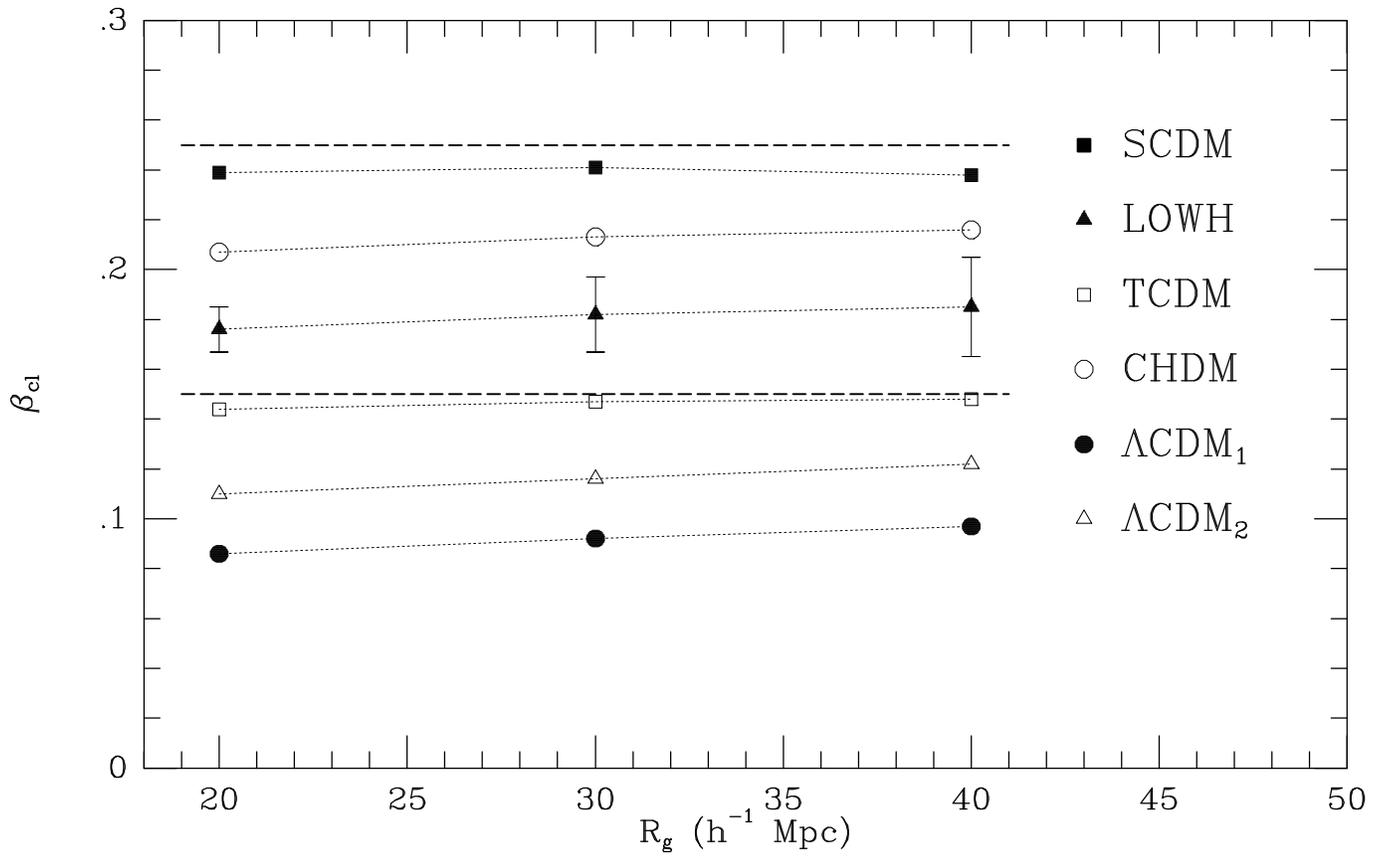